\documentclass[%
groupedaddress,
preprint,
amsmath,amssymb,
aps,
pre,
]{revtex4-2}

\usepackage{graphicx}
\usepackage{dcolumn}
\usepackage{bm}
\usepackage{mathtools}
\usepackage[normalem]{ulem}
\usepackage[utf8]{inputenc}
\usepackage[T1]{fontenc}
\usepackage{mathptmx}
\usepackage{etoolbox}
\usepackage{epsfig}
\usepackage{epstopdf}
\usepackage{subcaption}
\usepackage{latexsym}
\usepackage{float}
\usepackage{makecell}
\usepackage{hyperref}
\usepackage{xcolor}
\usepackage{booktabs}
\usepackage{multirow}
\usepackage{natbib}

\newcommand{\RN}[1]{%
  \textup{\lowercase\expandafter{\romannumeral#1}}%
}

\usepackage{todonotes}

\usepackage{epsfig}
\usepackage{amsfonts}
\usepackage{float}
\usepackage{amsthm}
\usepackage{tikz}
\usepackage{graphics,xcolor}
\usepackage[normalem]{ulem}
\usepackage{graphicx}
\usepackage{amssymb}
\usepackage{amsmath}
\usepackage{bigints}
\usepackage{amsfonts}
\usepackage[autostyle]{csquotes}
\usepackage{hyperref}
\usepackage{wrapfig}
\usepackage{url}
\usepackage{wasysym}

\usepackage{epstopdf}
\usepackage{epsfig}
\usepackage{caption}
\usepackage{subcaption}
\usepackage{wrapfig}
\usepackage{array,multirow}
\usepackage{caption}

\theoremstyle{definition}

\catcode`@=12

\makeatother

\begin{document}


\title{Towards enhanced mixing of a high viscous miscible blob in porous media}
\author{Mijanur Rahaman}
 \affiliation{Department of Mathematics, Indian Institute of Technology Guwahati, Guwahati 781039, India}%

\author{Jiten C. Kalita}
\affiliation{Department of Mathematics, Indian Institute of Technology Guwahati, Guwahati 781039, India}%

\author{Satyajit Pramanik}
\email{satyajitp@iitg.ac.in}
\affiliation{Department of Mathematics, Indian Institute of Technology Guwahati, Guwahati 781039, India}%

\date{\today}

\begin{abstract}
In this study, we investigate the rectilinear displacement and deformation of a highly viscous, miscible circular blob influenced by a less viscous fluid within a homogeneous porous medium featuring physically realistic no-flux boundaries. We utilize a fourth-order accurate compact finite difference scheme for the spatial discretization of the nonlinear partial differential equations that govern this phenomenon. The resulting semi-discrete equations are then integrated using the second-order Crank-Nicolson (CN) method. We conduct numerical simulations for a P\'eclet number ($Pe \leq 3000$) and a log-mobility ratio $0 \leq R \leq 7$, which reveal three distinct pattern formations: comet-shape, lump-shape, and viscous fingering instability. Our results demonstrate that the deformation, spreading, and mixing of the blob vary non-ideally with both $Pe$ and $R$, a behavior attributed to the blob's initial curvature. Consequently, enhanced mixing can be achieved at intermediate values of $Pe$ and $R$, suggesting the existence of an optimal mixing condition. These findings have significant implications for fields such as oil recovery, CO$_2$ sequestration, pollution remediation, and chromatography separation.

\end{abstract}

    \maketitle
    
\section{Introduction} \label{sec:intro}

Fluid flow and mixing in porous media are crucial areas of research, given their significance in various industrial and environmental processes, such as oil recovery \citep{homsy1987viscous}, carbon dioxide sequestration \citep{huppert2014fluid}, pollutant remediation, and chromatography separation \citep{de2005viscous}. The physico-chemical properties of the flow can lead to hydrodynamic instabilities in porous media, including Saffman-Taylor \citep{saffman1958penetration} and Rayleigh-Taylor instability \citep{riaz2006onset}, among others. When a less viscous fluid drives a more viscous fluid in a porous medium, the interface between the two is prone to instability, known as the Saffman-Taylor instability. This instability results in the formation of finger-like structures, wherein the upstream fluid invades the downstream fluid, a phenomenon referred to as viscous fingering (VF). Viscous fingering can occur with both miscible and immiscible fluids, with the instability in immiscible fluids first documented by \citet{saffman1958penetration}. 

To understand the mechanisms behind VF instability in porous media, researchers have conducted theoretical, experimental, and numerical studies on both miscible and immiscible systems using rectilinear and radial geometries since the early foundational works. In an immiscible fluid system, solubility is zero, while in a fully miscible system, it is infinite. Recently, there has been growing interest in investigating the effects of VF in partially miscible fluids. The current work focuses on the effects of VF caused by the introduction of a different viscous sample within a fully miscible system.

Miscible viscous fingering of a highly viscous sample displaced by a less viscous one has been the subject of both experimental \citep{maes2010experimental} and numerical \citep{de2005viscous, mishra2008differences, pramanik2016fingering} studies over the years. \citet{de2005viscous} investigated the fingering dynamics of a more viscous slice. Later, \citet{mishra2008differences} extended this work by considering cases where the sample could be either more or less viscous and comparing the effects in each scenario. They found that due to the flow direction, viscous fingering (VF) had a more pronounced effect on variance or mixing length in low-viscosity samples, where VF formed at the frontal interface. The current work is primarily inspired by the experimental findings of \citet{maes2010experimental}, which analyzed the dispersion and VF of an initially circular, more viscous sample within a rectilinear displacement in a horizontal Hele-Shaw cell. The unstable viscosity contrast between the displacing and displaced fluids resulted in a finger-shaped pattern at the rear of the blob, while a downstream tail pattern was observed at the front. Additionally, the rectilinear displacement of a more viscous circular blob by a less viscous fluid was numerically investigated by \citet{pramanik2015viscous} using the pseudo-spectral method \citep{tan1988simulation}. They reported three distinct pattern formations depending on the viscosity contrast parameter $(R)$, P\'eclet number $(Pe)$, and the radius $(r)$ of the blob, and noted a parameter-dependent critical window for VF. Later, in another study, \citet{pramanik2016fingering} explored the fingering dynamics and deformation of a less viscous blob. They showed that the critical radius for inducing VF is smaller for a less viscous blob than for a more viscous one, and that the concept of a critical window for VF does not apply to less viscous blobs. Their simulations were limited by numerical instability at certain parameter values. 
The objective of the current study is to explore the mechanism of VF instability in a finite high-viscosity fluid domain sandwiched between lower-viscosity fluid domains across a broader range of parameter space than what has been documented in the literature, thereby extending beyond existing experimental limits. 

In addition to the previously mentioned Fourier pseudo-spectral method, numerous well-established numerical techniques are available in the literature for solving miscible viscous fingering problems. A higher-order discontinuous Galerkin (DG) method with weighted average stabilization and flux reconstruction has been employed to simulate VF problems related to miscible displacement in porous media \citep{li2015high, li2016numerical}. Compared to the cell-centered finite volume approach, this method offers superior computational efficiency and significantly reduced sensitivity to grid orientation \citep{li2016numerical}. In another study \citep{hidalgo2012scaling}, the streamfunction-vorticity approach was utilized for simulation, where a spectral Galerkin method was used for the stream function, and a compact sixth-order finite difference scheme was applied for concentration. Additionally, \citet{islam2005fully} implemented an alternating-direction implicit (ADI) technique combined with a Hartley-based pseudo-spectral method for similar computations. While the pseudo-spectral method is known for its accuracy, it necessitates periodic boundary conditions. A higher-order compact (HOC) finite difference method offers a way to overcome the restriction of boundary conditions without compromising accuracy, making it the preferred choice for simulating the phenomenon. Researchers have employed various strategies to achieve higher-order compactness in finite difference approximations. For instance, \citet{gupta1984single} used a series expansion of the differential equations, while \citet{dennis1989compact} applied a transformation involving the exponential expansion of a definite integral of the convective coefficient of the relevant partial differential equation. Other notable works on HOC schemes include the discrete-weighted mean approximation methods by \citep{gartland1982discrete, noye1988third, noye1989finite}, the weighted modified PDE method by \citep{tang1995compact}, and high-order upwind schemes by \citep{wilkes1983evaluation, yanwen1999numerical, sesterhenn2000characteristic}.

Another way of achieving higher-order compactness is by using the original differential equation to substitute for the leading truncation error terms of the standard central difference approximation. This idea was first implemented by Lax and Wendroff \cite{lax1959systems} on the time-dependent hyperbolic partial differential equations (\cite{pletcher2012computational, anderson2002computational}). 
They used this technique to increase the temporal accuracy; however, the spatial implementation of this temporal Lax-Wendroff idea was first proposed by \citet{mackinnon1988analysis}. Later on, \citet{spotz1994high, spotz1995high}  extended this scheme to solve the 2D steady-state Navier-Stokes (N-S) equation. This idea was further expanded by \citet{kalita2002class}, who developed a HOC schemes for the two-dimensional unsteady convection-diffusion equation with variable coefficients, making them suitable for solving the unsteady  N-S equations.

In this work, we adopt a temporally second-order and spatially fourth-order accurate HOC finite difference method \cite{kalita2002class, kalita2001hoc} on a uniform grid to numerically investigate the rectilinear displacement of a miscible circular blob displaced by a less viscous fluid. This reconstructed scheme enables simulations over a broader range of parameter values and allows for subsequent analysis. We utilize the streamfunction-vorticity formulation of the governing equations, which includes a convection-diffusion equation for solute concentration coupled with Darcy's law for fluid velocity, while adhering to the incompressibility condition of the fluid. 


The outline of this article is as follows. Section \ref{sec:formulation} describes the problem formulation, numerical procedure, and validation study. In \S \ref{sec:results}, we present the findings of our study and discuss the obtained results, followed by concluding remarks in \S \ref{sec:conclusion}. 

\section{Mathematical formulation and numerical solution} \label{sec:formulation}

\begin{figure}[htbp]
\centering
    \includegraphics[width = 0.8\textwidth]{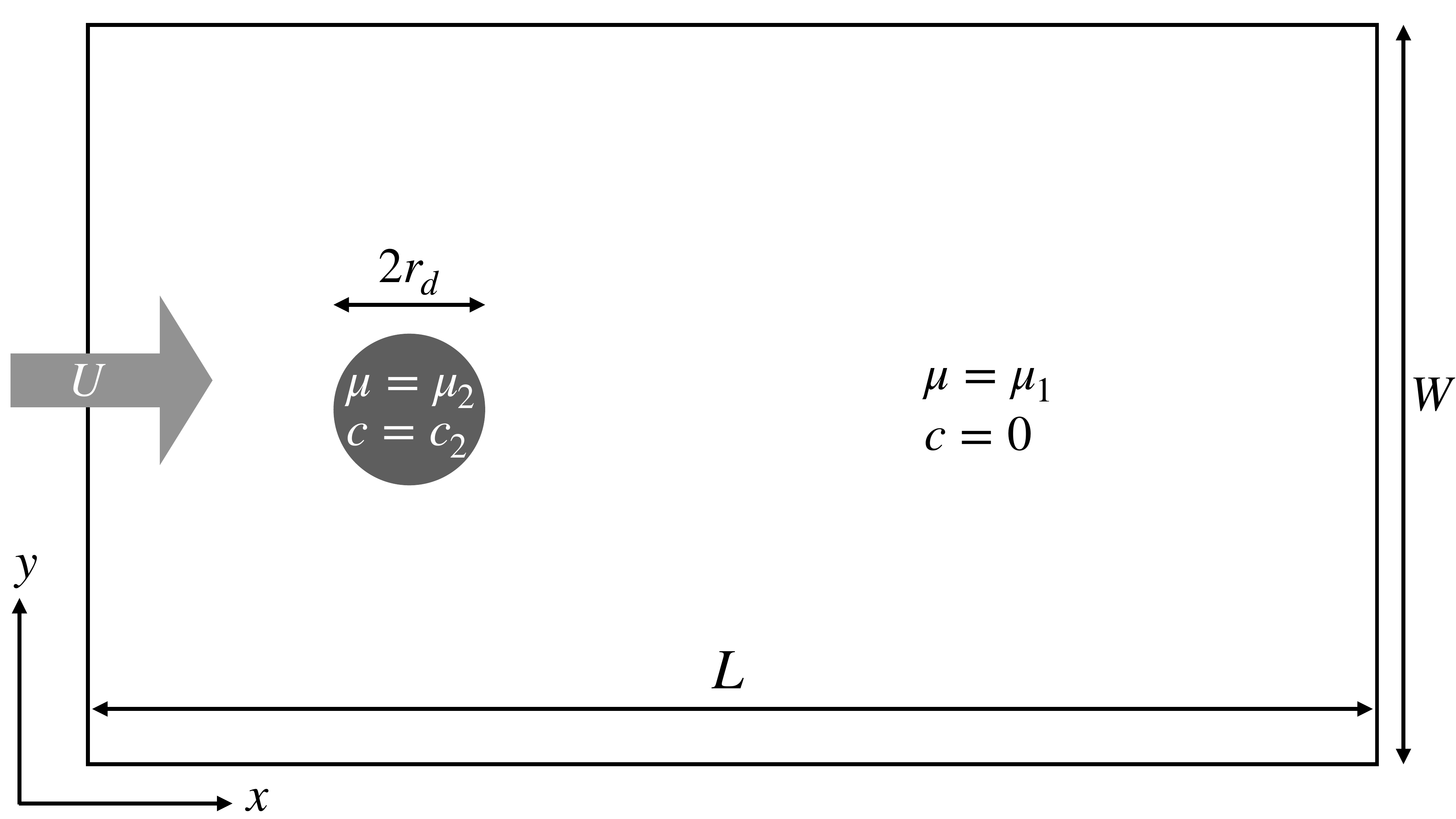}
    \caption{Schematic of the displacement of a circular blob in a two-dimensional homogeneous porous medium (not to be scaled).} 
\label{fig:schematic}
\end{figure}

Consider the rectilinear displacement of a miscible circular blob in a two-dimensional homogeneous porous medium of dimension $L \times W$. The circular blob of radius $r_d$, and viscosity $\mu_2$ contains a solute of concentration $c = c_2$; whereas, the solute is initially absent from the displacing fluids ($c = 0$) that is injected at a constant speed $U$ along the $x$-direction (see figure \ref{fig:schematic} for a schematic). Fluids are assumed to be incompressible, neutrally buoyant, and non-reactive; the viscosity depends on the solute concentration. 


\subsection{Governing equations and non-dimensional formulation} \label{subsec:non-dimensional}

We assume a porous medium with low porosity such that the porescale Reynolds number is negligible and hence the flow is governed by non-inertial Darcy's law, 
\begin{equation}
    \label{eq:darcy_law}
    \boldsymbol{\nabla}\cdot\boldsymbol{u} = 0, \qquad \boldsymbol{u} = -\frac{\kappa}{\mu(c)}\boldsymbol\nabla p, 
\end{equation}
where $p$ is the hydrodynamic pressure, $\boldsymbol{u} = (u, v)$ is the two-dimensional velocity vector, $\kappa$ is the permeability of the porous medium, $\mu$ is the dynamic viscosity of the fluid. Assuming a homogeneous, isotropic dispersion of the solute in the solvent, transport of the solute can be described via an advection-diffusion equation
\begin{equation}
    \label{eq:transport}
    \frac{\partial c}{\partial t} + \boldsymbol \nabla \cdot (\boldsymbol{u} c) = D \nabla^2 c, 
\end{equation}
where the isotropic dispersion coefficient $D$ is assumed to be constant. 

We render the non-dimensional variables
\begin{equation}
    \label{eq:scaling}
    (x^\prime, y^\prime) = \frac{(x, y)}{L_c}, \;\; \boldsymbol{u}^\prime = \frac{\boldsymbol{u}}{U}, \;\; t^\prime = \frac{t}{L_c/U}, \;\; p^\prime = \frac{p}{\mu_1 U L_c/\kappa}, \;\; \mu^\prime = \frac{\mu}{\mu_1}, \;\; c^\prime = \frac{c}{c_2}, 
\end{equation}
where the characteristic length scale $L_c = W/16$. 
The corresponding dimensionless equations, in a reference frame moving with the injection velocity, read (after dropping the prime symbols) 
\begin{eqnarray}
\label{eq:nondim_incompressibility}
\boldsymbol\nabla\cdot\boldsymbol{u} & = & 0,  \\
\label{eq:nondim_Darcy}
\boldsymbol\nabla p & = & -\mu(c)(\boldsymbol{u} + \boldsymbol{i}), \\ 
\label{eq:nondim_transport}
\frac{\partial c}{\partial t} + \boldsymbol\nabla\cdot(\boldsymbol{u}c) & = & \frac{1}{Pe}\nabla^2c,
\end{eqnarray}
where $\boldsymbol{i}$ is the unit vector in $x$-direction and the concentration-dependent viscosity is assumed to follow Arrhenius relation \citep[see][and reference therein]{pramanik2015viscous}
\begin{equation}
    \label{eq:arrhenius}
    \mu(c) = e^{Rc} \quad. 
\end{equation}
A close look into equations \eqref{eq:nondim_incompressibility}-\eqref{eq:arrhenius} yields that miscible displacement in the porous media can be studied in terms of two dimensionless parameters --- P\'eclet number ($Pe$) and log-mobility ratio ($R$) defined as
\begin{equation}
    Pe = \frac{U L_c}{D} \qquad \mbox{and} \qquad R = \ln\left( \frac{\mu_2}{\mu_1} \right). 
\end{equation}
Introducing stream function $\psi(x,y,t)$, such that $\displaystyle \boldsymbol{u} = \left( \frac{\partial \psi}{\partial y}, -\frac{\partial \psi}{\partial x} \right)$ and by eliminating the pressure term from the governing equations, equations \eqref{eq:nondim_incompressibility}-\eqref{eq:nondim_transport} reduce to  
\begin{eqnarray}
    \label{eq:SF1}
    \nabla^2\psi = -R\boldsymbol \nabla c \cdot{(\boldsymbol{\nabla}\psi + \boldsymbol{j})}, \\ 
    \label{eq:SF2}
    \frac{\partial c}{\partial t} + \frac{\partial \psi}{\partial y}\frac{\partial c}{\partial x} - \frac{\partial \psi}{\partial x}\frac{\partial c}{\partial y} = \frac{1}{Pe} \nabla^2c, 
\end{eqnarray}
where $\boldsymbol{j}$ is the unit vector in $y$-direction. 

\subsection{Initial and boundary conditions} \label{subsec:initial_boundary_conditions}

A description of appropriate initial and boundary conditions makes the mathematical formulation of the above problem complete. Initial conditions for concentration and stream function are 
\begin{eqnarray}
\label{eq:IC1}
& & c(x,y,0) = \begin{cases}
                1, \quad  x^2+y^2<r^2 \\
                0, \quad  \text{ elsewhere }
                \end{cases}, \\ 
\label{eq:IC2}
& & \boldsymbol{u}(x,y,0) = \left( \frac{\partial \psi}{\partial y}, -\frac{\partial \psi}{\partial x} \right) = (0,0),
\end{eqnarray}
where $r$ is the non-dimensional radius of the blob. Here, we consider no-flux boundaries for the solute concentration, 
\begin{equation}
    \label{eq:BC1}
    \boldsymbol{\nabla}{c} \cdot \boldsymbol{n} = 0. 
\end{equation}
For the stream function, at the transverse boundaries we have used
\begin{equation}
    \label{eq:BC2}
    \psi{(x,0,t)} = \psi{(x,L_y,t)} = 0, 
\end{equation}
and no-flux conditions at the longitudinal boundaries
\begin{equation}
    \label{eq:BC3}
    \left.\frac{\partial \psi}{\partial x}\right|_{(0,y,t)} = \left.\frac{\partial \psi}{\partial x}\right|_{(L_x,y,t)} = 0. 
\end{equation}
Here, $L_x$ and $L_y$ are the dimensionless length and width of the computational domain, and $\boldsymbol{n}$ is the outward unit normal vector. 


\subsection{Fourth-order accurate compact finite difference scheme} \label{subsec:numerical_scheme} 

The inherent nonlinearities of the coupled advection-diffusion system \eqref{eq:SF1}-\eqref{eq:SF2} incorporate challenges to obtain analytical solutions. Therefore, we resort to numerical solutions by discretizing the equations using a higher-order compact scheme  that has been widely used in the studies of advection-diffusion problems \citep{kalita2002class} having spatial accuracy four. 
 We use a Crank-Nicolson type of discretization for the time derivative, ensuring a second-order temporal accuracy. The computational domain $\displaystyle \left[ -2L_x/3, L_x/3 \right] \times \left[ -L_y/2, L_y/2 \right] \times \left[ 0, T \right]$ is discretized using a uniform mesh of size $N_x \times N_y \times N_t$, where $N_x$, $N_y$, and $N_t$ are the number of grids taken along the $x$-, $y$-, and $t$-direction, respectively. The dependent variable $\phi$ at $(x_i, y_j, t_n)$ is denoted as $\phi_{i,j}^{(n)}$, where $x_i = -2L_x/3 + ih, \; y_j = -L_y/2 + jh$ and $t_n = n\Delta t$ for $i = 0, 1, \hdots, N_x-1$, $j = 0, 1, \hdots, N_y-1$ and $n = 0, 1, \hdots, N_t-1$ with $h = L_x/(N_x-1) = L_y/(N_y-1)$ and $\Delta t = T/(N_t-1)$. 


The HOC scheme approximation of the equation \eqref{eq:SF2} is obtained as
\begin{eqnarray}
\label{eq:HOC_generalized} 
& & Pe \left[ 1 + \frac{h^{2}}{12} \left( \delta_{x}^{2}+\delta_{y}^{2}-b_{i j} \delta_{x}-d_{i j} \delta_{y} \right) \right] \left(c_{ij}^{(n+1)}-c_{ij}^{(n)} \right) \nonumber \\ 
& & \qquad =\frac{\Delta t}{2} \left[ \alpha_{i j} \delta_{x}^{2} + \beta_{i j} \delta_{y}^{2}-B_{i j} \delta_{x}-D_{i j} \delta_{y} + \frac{h^{2}}{6} \left( \delta_{x}^{2} \delta_{y}^{2}-b_{i j} \delta_{x} \delta_{y}^{2}-d_{i j} \delta_{x}^{2} \delta_{y}-\gamma_{i j} \delta_{x} \delta_{y} \right) \right] \left( c_{ij}^{(n+1)}+c_{ij}^{(n)} \right) \nonumber \\ 
& & \qquad \qquad \qquad \qquad \qquad \qquad \qquad \qquad \qquad \qquad + \mathcal{O} (\Delta t^2, h^4),
\end{eqnarray}
where $\delta_{x}, \; \delta_{y}, \; \delta_{x}^{2}, \; \delta_{y}^{2}$ are the first and second order central difference operators, $h$ is uniform step length along $x$ and $y$ directions respectively, and $\Delta t$ is the uniform time step length. The coefficients  $b_{ij},d_{ij},B_{ij},D_{ij},\alpha_{i j},\beta_{i j}$ and $\gamma_{i j}$ are defined as follows:
\begin{eqnarray}
\label{eq:bij}
& & b_{ij} = \left( Pe\frac{\partial \psi}{\partial y} \right)_{ij}^{(n)}, \\
\label{eq:dij}
& & d_{ij} = -\left( Pe\frac{\partial \psi}{\partial x} \right)_{ij}^{(n)}, \\
\label{eq:Bij}
& & B_{ij} = \left[ 1 + \frac{h^{2}}{12} \left( \delta_{x}^{2} + \delta_{y}^{2}-b_{i j} \delta_{x}-d_{i j} \delta_{y} \right) \right] b_{i j}, \\
\label{eq:Dij}
& & D_{ij} = \left[ 1 + \frac{h^{2}}{12} \left( \delta_{x}^{2} + \delta_{y}^{2}-b_{i j} \delta_{x} - d_{i j} \delta_{y} \right) \right] d_{i j}, \\
\label{eq:alphaij}
& & \alpha_{ij} = 1 + \frac{h^{2}}{12} \left( b_{i j}^{2}-2 \delta_{x} b_{i j} \right), \\
\label{eq:betaij}
& & \beta_{ij} = 1 + \frac{h^{2}}{12} \left( d_{ij}^{2}-2 \delta_{y} d_{i j} \right), \\
\label{eq:gammaij}
& & \gamma_{ij} = \delta_{x} d_{i j} + \delta_{y} b_{i j}-b_{i j} d_{i j}.
\end{eqnarray}

Equation \eqref{eq:HOC_generalized} can be expressed in the matrix form as
\begin{equation}
\label{eq:matrix_system1}
    \mathbb{A} \boldsymbol{c}^{(n+1)} = \boldsymbol{f}(\boldsymbol{c}^{(n)}),
\end{equation}
where the coefficient matrix $\mathbb{A}$ is an square matrix of order $N_x N_y$, and $\boldsymbol{c}^{n+1}$ is the unknown concentration vector at the $(n+1)^{\rm th}$ time level having $N_x N_y$ components. 

Once the concentration values are computed at the current time level, the streamfunction $\psi$ is computed from equation \eqref{eq:SF1} via the steady-state form of the schemes \citep[see][and references therein for further details]{kalita2002class, kalita2001hoc}. Thus, the HOC approximation to equation \eqref{eq:SF1} yields
\begin{eqnarray}
\label{eq:HOC_generalized2}
& & -\tilde{\alpha}_{i j}\delta_{x}^{2} \psi_{i j} - \tilde{\beta}_{i j}\delta_{y}^{2} \psi_{i j} + \tilde{B}_{i j} \delta_{x} \psi_{i j} + \tilde{D}_{i j} \delta_{y} \psi_{i j} - \frac{h^{2}}{6}\left[\delta_{x}^{2} \delta_{y}^{2}-\tilde{b}_{i j} \delta_{x} \delta_{y}^{2}-\tilde{d}_{i j} \delta_{x}^{2} \delta_{y}-\tilde{\gamma}_{i j} \delta_{x} \delta_{y}\right] \psi_{i j} \nonumber \\ 
& & \qquad \qquad \qquad \qquad \qquad \qquad \qquad \qquad \qquad \qquad \qquad \qquad =\tilde{F}_{i j} + \mathcal{O}(h^4),
\end{eqnarray}
where the coefficients  $\tilde{b}_{ij}, \; \tilde{d}_{ij}, \; \tilde{B}_{ij}, \; \tilde{D}_{ij}, \; \tilde{\alpha}_{i j}, \; \tilde{\beta}_{i j}$, $\tilde{\gamma}_{i j}$ and $\tilde{F}_{i j}$ are defined as 
\begin{eqnarray}
\label{eq:bij_tilde}
& & \tilde{b}_{ij} = -R\delta_{x}c_{ij}^{(n+1)}, \\
\label{eq:dij_tilde}
& & \tilde{d}_{ij} = -R\delta_{y}c_{ij}^{(n+1)}, \\
\label{eq:Bij_tilde}
& & \tilde{B}_{ij} = \left[ 1 + \frac{h^{2}}{12} \left( \delta_{x}^{2} + \delta_{y}^{2} - \tilde{b}_{i j} \delta_{x} - \tilde{d}_{i j} \delta_{y} \right)  \right] \tilde{b}_{i j}, \\
\label{eq:Dij_tilde}
& & \tilde{D}_{i j} = \left[ 1 + \frac{h^{2}}{12} \left( \delta_{x}^{2} + \delta_{y}^{2} - \tilde{b}_{i j} \delta_{x} - \tilde{d}_{i j} \delta_{y} \right) \right] \tilde{d}_{i j}, \\
\label{eq:Fij_tilde}
& & \tilde{F}_{i j} = R \left[ 1 + \frac{h^{2}}{12} \left( \delta_{x}^{2} + \delta_{y}^{2} - \tilde{b}_{i j} \delta_{x} - \tilde{d}_{i j} \delta_{y} \right) \right] \delta_{y}c_{i j}^{(n+1)}, \\ 
\label{eq:alphaij_tilde}
& & \tilde\alpha_{i j} = 1 + \frac{h^{2}}{12} \left( \tilde{b}_{i j}^{2} - 2 \delta_{x} \tilde{b}_{i j} \right), \\
\label{eq:betaij_tilde}
& & \tilde{\beta_{i j}} = 1 + \frac{h^{2}}{12} \left( \tilde{d}_{i j}^{2} - 2 \delta_{y} \tilde{d}_{i j} \right), \\ 
\label{eq:gammaij_tilde}
& & \tilde{\gamma_{i j}} = \delta_{x} \tilde{d}_{i j} + \delta_{y} \tilde{b}_{i j} - \tilde{b}_{i j} \tilde{d}_{i j}. 
\end{eqnarray} 
Equation \eqref{eq:HOC_generalized2} can be written in matrix form as
\begin{equation}
    \label{eq:matrix_system2}
    \tilde{\mathbb{A}} \boldsymbol{\psi}^{(n+1)} = \tilde{\boldsymbol{f}}(\boldsymbol{c}^{(n+1)}), 
\end{equation}
where the coefficient matrix $\tilde{\mathbb{A}}$ is an square matrix of order $N_x N_y$, and $\boldsymbol{\psi}^{(n+1)}$ is the unknown stream function vector at $(n+1)^{\rm th}$ time level having $N_x N_y$ components. 

Once $\boldsymbol{\psi}^{(n+1)}$ is available, the velocities at the $(n+1)^{th}$ time level are computed by using a fourth-order approximation based on the Pad\'e approach \citep{gupta2005new}, 
\begin{eqnarray}
    \label{eq:u_velocity}
    \frac{1}{4}u_{i, j-1}^{(n+1)} + u_{i, j}^{(n+1)} + \frac{1}{4}u_{i, j+1}^{(n+1)} = \frac{3}{4h} \left( \psi_{i, j+1}^{(n+1)} - \psi_{i, j-1}^{(n+1)} \right), 
\end{eqnarray}
and
\begin{eqnarray}
    \label{eq:v_velocity}
    \frac{1}{4}v_{i-1, j}^{(n+1)}+v_{i, j}^{(n+1)} + \frac{1}{4}v_{i+1,j}^{(n+1)} = -\frac{3}{4h} \left( \psi_{i+1, j}^{(n+1)} - \psi_{i-1, j}^{(n+1)} \right). 
\end{eqnarray}
The tri-diagonal system of linear equations, \eqref{eq:u_velocity} and \eqref{eq:v_velocity}, is solved using Thomas' algorithm. 


The time marching of the coupled system is achieved incorporating an inner and outer iterative procedure. The inner iterations are composed of solving the matrix equations \eqref{eq:matrix_system1} and \eqref{eq:matrix_system2} using the biconjugate gradient stabilized (BiCGStab) iterative solver. An incomplete LU decomposition is used as a preconditioner with the help of the Lis library \cite{lisnew}. The iterations are stopped when the norm of the residual vectors arising out of the respective system of equations falls below the tolerance limit $10^{-7}$. Computations of $c$, $\psi$, $u$, and $v$ employing equations \eqref{eq:matrix_system1}, \eqref{eq:matrix_system2}--\eqref{eq:v_velocity}, constitute one outer iteration. The same is repeated at each time level until the desired final time is reached. 

\subsection{Grid independence and validation} \label{subsec:validation}

To ensure that the boundary does not affect the dynamics of the blob, we choose a large domain and keep the blob far away from the boundary such that $L_x: L_y = 3: 2, \; L_y/r\geq 32$. Thus, the width and length of the non-dimensional computational domain are $L_y = 16$ and $L_x = 24$, respectively; the non-dimensional radius of the blob is $r \leq 0.5$. All the results reported in this paper are based in numerical simulations performed for $T = 10$ unless mentioned otherwise. Rigorous validation and grid independence tests are executed for the proposed HOC scheme as discussed below. 

As a benchmark for validating the developed numerical scheme, we consider the displacement of a viscosity-matched miscible blob ($R = 0$) for different P\'eclet numbers in the range $500 \leq Pe \leq 3000$. In the absence of viscosity contrast between the displacing and displaced fluids, the blob is carried along with the displacing fluid. Thus, in the moving frame of reference, the dynamics of the circular blob can be described by solving a 2D diffusion equation subjected to no-flux boundary conditions and an appropriate initial condition. 

To better understand the dynamics of miscible fluids, we resort to various qualitative and quantitative measures described below. The transverse-averaged concentration profile is defined as 
\begin{equation}
    \label{eq:trans_avg}
    \bar{c}(x,t)=\frac{1}{L_y}\int\limits_{0}^{L_y}c(x,y,t) {\rm d} y. 
\end{equation}
The first moment, also known as the center of mass of the distribution, is defined as 
\begin{equation}
    \label{eq:first_moment_x}
    m_x(t) = \frac{\int_{0}^{L_x}x\bar{c}(x,t) {\rm d} x}{\int_{0}^{L_x}\bar{c}(x,t) {\rm d} x}. 
\end{equation}
These measures help quantify the effects of diffusion in the longitudinal direction. On the other hand, to quantify the effects of diffusion in the transverse direction, we define the longitudinal-averaged concentration profile 
\begin{equation}
    \label{eq:long_avg}
    \tilde{c}(y,t) = \frac{1}{L_x} \int \limits_{0}^{L_x}c(x,y,t) {\rm d} x,
\end{equation}
and the corresponding first moment  
\begin{equation}
    \label{eq:first_moment_y}
    m_y(t) = \frac{\int_{0}^{L_y}y\tilde{c}(y,t) {\rm d} y}{\int_{0}^{L_y} \tilde{c}(y,t) {\rm d} y}. 
\end{equation}

\begin{figure}[!htbp]
    \centering
    (a) \hspace{3.2 in} (b) \\ 
    \includegraphics[width=0.495\textwidth]{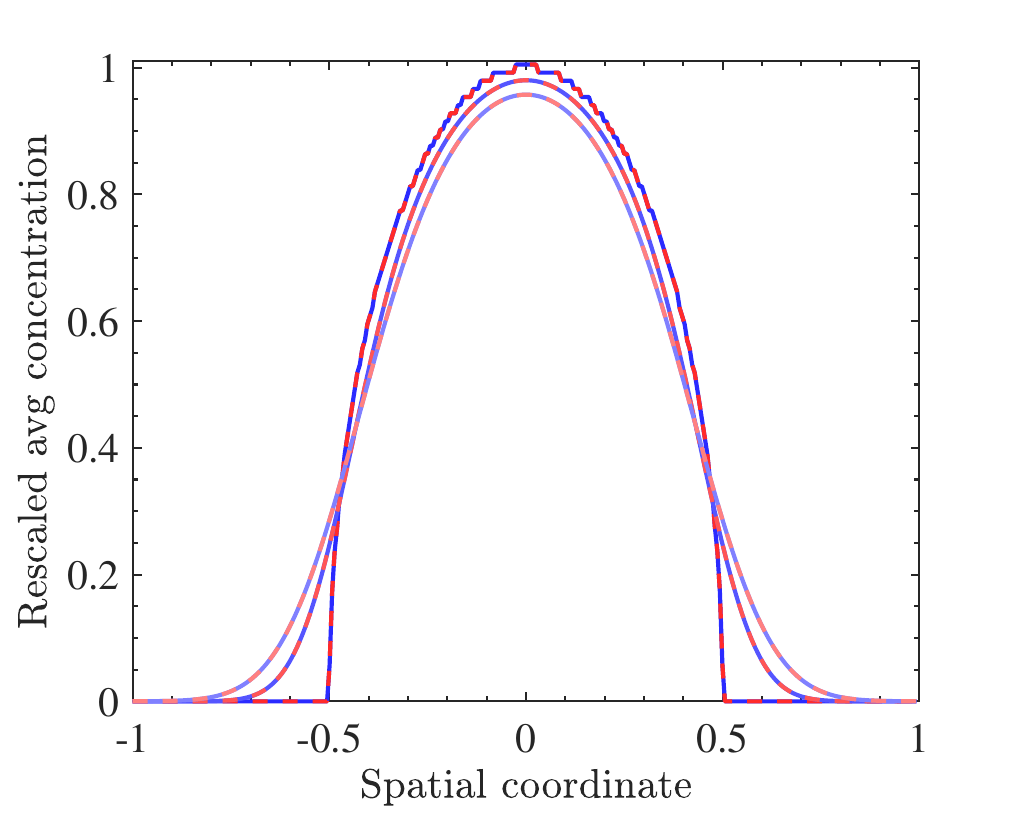} 
    \includegraphics[width=0.495\textwidth]{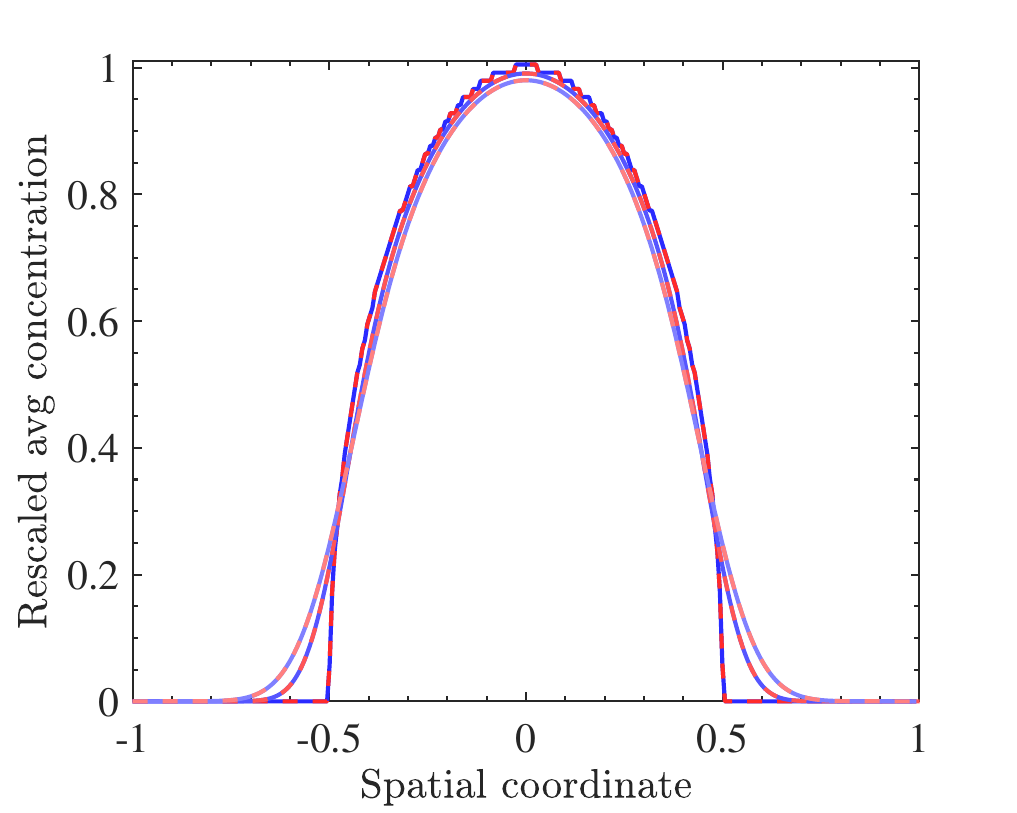} \\
    (c) \\
    \includegraphics[width=0.5\textwidth]{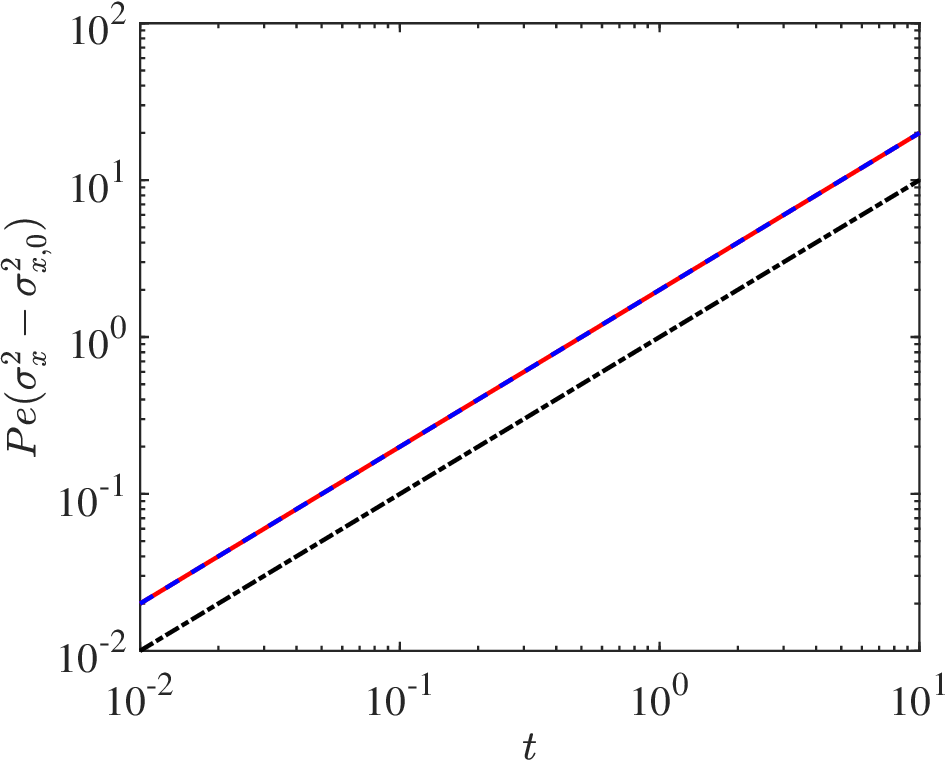} 
    \caption{Rescaled longitudinal- and transverse-averaged concentration, $L_y \bar{c}(x, t)$ (red curves) and $L_x \tilde{c}(y, t)$ (blue curves), respectively, at $t = 0, 5$ and $10$ (dark to light) (a) $Pe = 1000$, (b) $Pe = 2000$. These curves are indistinguishable due to the circular symmetry of the diffusive phenomenon. This ensures the robustness of the HOC scheme developed and used here. (c) Rescaled longitudinal variance, $Pe(\sigma_x^2(t) - \sigma_{x, 0}^2)$, for $Pe = 1000$ (red) and $Pe = 2000$ (blue), grows linearly in time. A black dash-dotted line is shown for reference.}
    \label{fig:R0_validation}
\end{figure}

As anticipated, the solute blob diffuses isotropically and mixes with the displacing fluid. Due to the radial symmetry of the diffusive mixing, the problem can be described in terms of one-dimensional diffusion, characterized by the variance varying linearly with time. Figure \ref{fig:R0_validation}(a-b) demonstrates, for each $Pe$, the spatial distributions of the rescaled longitudinal- and transverse-averaged concentration, $L_y \bar{c}(x, t)$ and $L_x \tilde{c}(y, t)$, respectively, are indistinguishable at all time. Furthermore, figure \ref{fig:R0_validation}(c) shows that the longitudinal variance of the solute evolves as $(\sigma_x^2 - \sigma_{x, 0}^2) \propto t/Pe$. These establish the robustness of the numerical method developed and used in this paper. 


\begin{figure}[!htbp]
    \centering
    \includegraphics[width=0.37\textwidth]{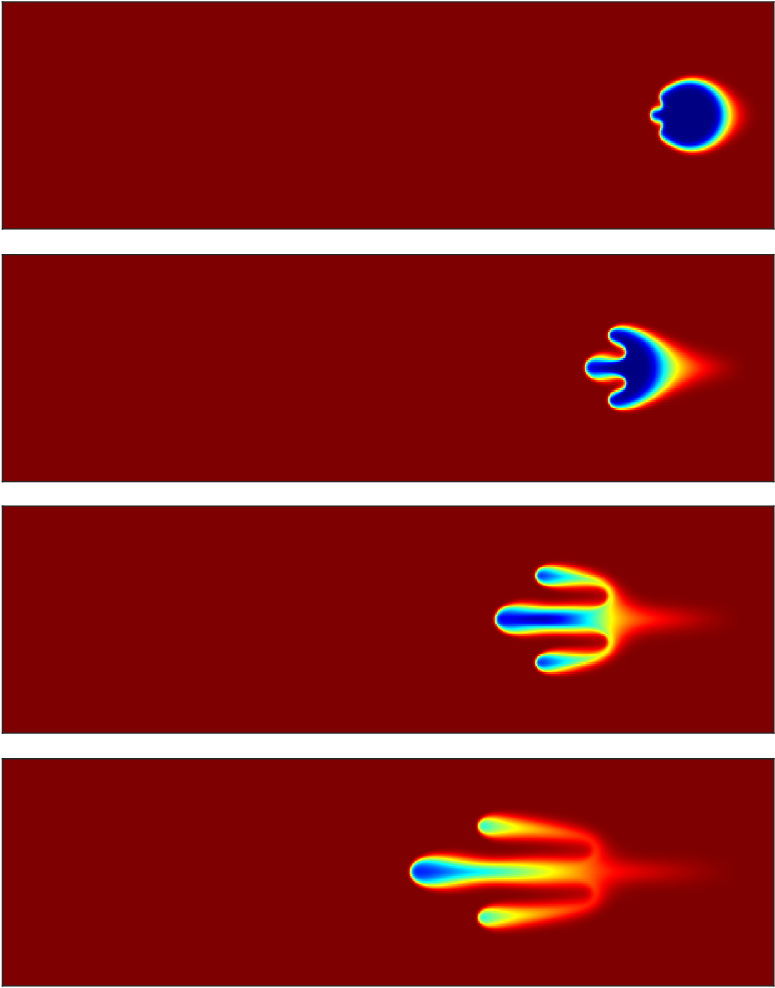} 
    \includegraphics[width=0.59\textwidth]{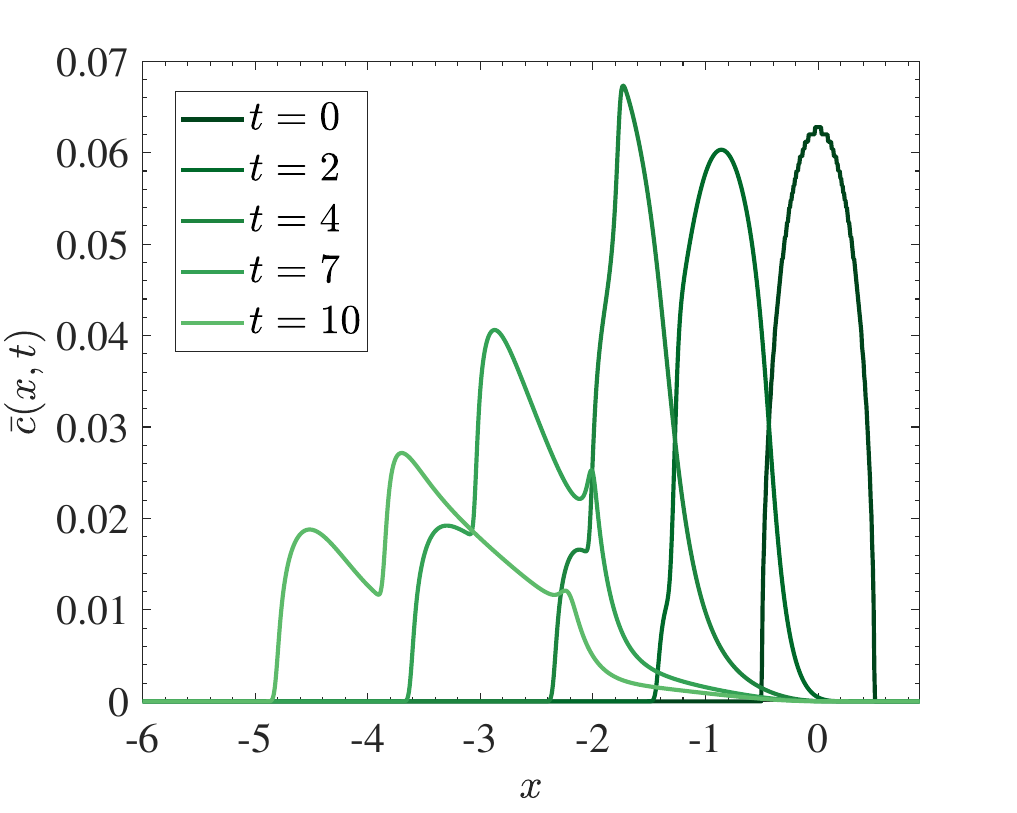} 
    \caption{Spatio-temporal distribution of a circular blob at different times, $t = 2, 4, 7, 10$ (from top to bottom), depicts fingering dynamics symmetric about the centerline (left panel). Corresponding transverse-average concentration plots (including the initial concentration) capture the fingering dynamics characterized by the multi-modal distribution with a downstream tail (right panel). The parameter values are: radius of the blob, $r = 0.5$, P\'eclet number, $Pe = 1000$, and log-mobility ratio, $R = 1$. These results are in excellent agreement with \citet{pramanik2015viscous}. Present HOC simulations are performed using $N_x = 3751$, $N_y = 2501$ and $\Delta t = 10^{-2}$.}
    \label{fig:R1_Pe1000_validation}
\end{figure}

\begin{figure}[!htbp]
    \centering
    \includegraphics[width=0.8\textwidth]{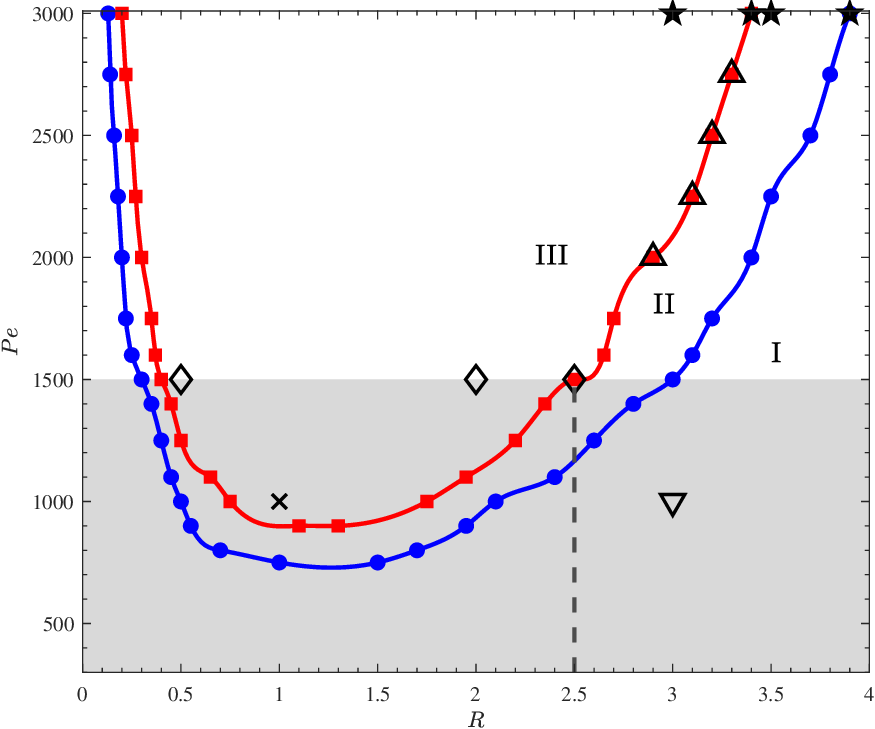}
    \caption{Phase plot in the $R-Pe$ plane for $r = 0.5$ featuring three distinct instability zones: Comet shape (I), lump shape (II), and VF (III). Markers (filled squares and filled circles)
    represent the parameter values used to approximate the boundaries between two adjacent zones. The grey shaded region marks the range of $Pe$, while the dashed vertical line corresponds to the upper limit of $R$ reported in the phase-plane analysis of \citet{pramanik2015viscous}.} 
    \label{fig:Phase_plane}
\end{figure}

Next, we consider the displacement of a highly viscous circular blob represented by $R > 0$ for which the viscosity at the rear semi-circular interface between the displacing and the displaced fluid is unstably stratified. We performed a rigorous grid independence study to investigate the deformation and viscous fingering of a circular blob of radius $r = 0.5$ for different values of $R$ and $Pe$. We used three different spatial grid sizes $1501 \times 1001$, $1876 \times 1251$, and $3751 \times 2501$ along with temporal step size $\Delta t = 10^{-2}$. It is observed that increasing the grid size beyond $3751 \times 2501$ does not alter the pattern formation. For further validation, we calculated the total mass of the blob corresponding to all the grids under consideration, yielding a maximum relative error $0.01\%$ for the finest grid considered here. Therefore, all the computations shown in this paper are performed using the spatial grid size $3751 \times 2501$ and temporal step size $\Delta t = 10^{-2}$. For $R = 1$ and $Pe = 1000$ ($\times$ in figure \ref{fig:Phase_plane}), the dynamics of a circular blob of radius $r = 0.5$ depicted in figure \ref{fig:R1_Pe1000_validation} exhibited an excellent agreement with the literature (see figure 3b of \citet{pramanik2015viscous}). 


It is worth mentioning that numerical simulations in \citet{pramanik2015viscous} were carried out using periodic boundaries only, as it was required for the Fourier pseudospectral method. In order to understand whether the dynamics of the circular blob depend on the nature of the boundary conditions, we performed numerical simulations with periodic transverse boundaries, too. No discernible difference was seen in the viscous fingering dynamics (not shown due to brevity). In summary, besides being able to capture the deformation and fingering dynamics of a circular blob with periodic boundaries \citep{pramanik2015viscous}, the HOC scheme implemented in this work captured the fingering dynamics with no-flux boundaries, which were not investigated earlier in the literature. Moreover, the present HOC scheme combined with an implicit CN method allowed us to take a time step $\Delta t = 10^{-2}$, which is two orders of magnitude smaller than the time step used in the explicit Adams-Bashforth (predictor) and trapezoidal (corrector) integrator used in \citet{pramanik2015viscous}. 


\section{Results and discussion} \label{sec:results}

In this section, we investigate the effects of the log-mobility ratio ($R$) and P\'eclet number ($Pe$) on the dynamics of a miscible circular blob of radius $r = 0.5$. The deformation and viscous fingering dynamics are quantified in terms of several matrices, such as longitudinal- and transverse-averaged concentration profiles, their mean, variances, and degree of mixing. 


\subsection{Phase diagram} \label{subsec:phase_diagram}

We performed over 170 numerical simulations for different combinations of $Pe$ and $R$, ranging $500 \leq Pe \leq 3000$ and $0 \leq R \leq 7$. Numerical simulations revealed three different pattern formations -- comet shape (I), lump shape (II), and viscous fingering (III), summarized in the $R-Pe$ phase plane (see figure \ref{fig:Phase_plane}). 
Furthermore, we validated the critical values of $R$ and $Pe$ for $r = 0.5$ to be $R_c \approx 1.2$ and $Pe_c \approx 900$ in close agreement with \citet{pramanik2015viscous}. This phase plane shows that the region of viscous fingering pattern formation (II) expands with $Pe$. For example, the interval of fingering instability increased from $0.75 \leq R \leq 1.75$ to $0.5 \leq R \leq 2.2$ when $Pe$ increased from $1000$ to $1250$, again in good agreement with \citet{pramanik2015viscous}. 
Further increasing $Pe$, we captured the interval of fingering instability $0.2 \leq R \leq 3.4$ for $Pe = 3000$. This region of fingering instability is surrounded by a region in which the more viscous circular blob deformed into a lump shape. 

\begin{figure}[!htbp]
    \centering
    (a) \hspace{1.3 in} (b) \hspace{1.3 in} (c) \hspace{1.3 in} (d) \\ 
    \includegraphics[width=0.24\textwidth]{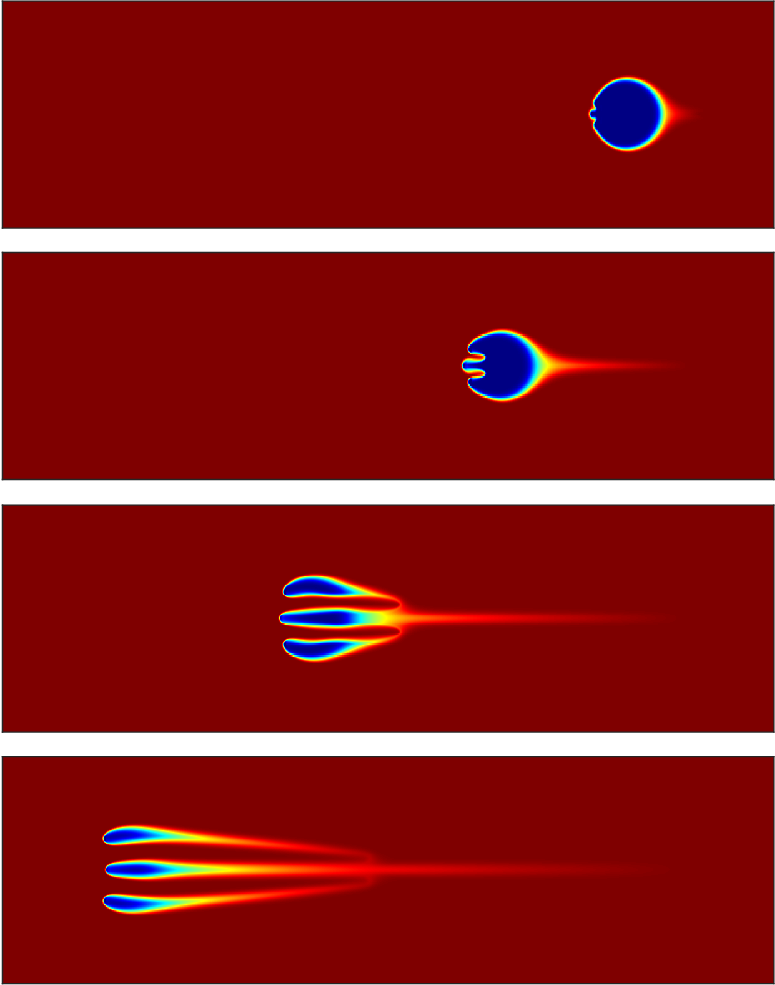}
    \includegraphics[width=0.24\textwidth]{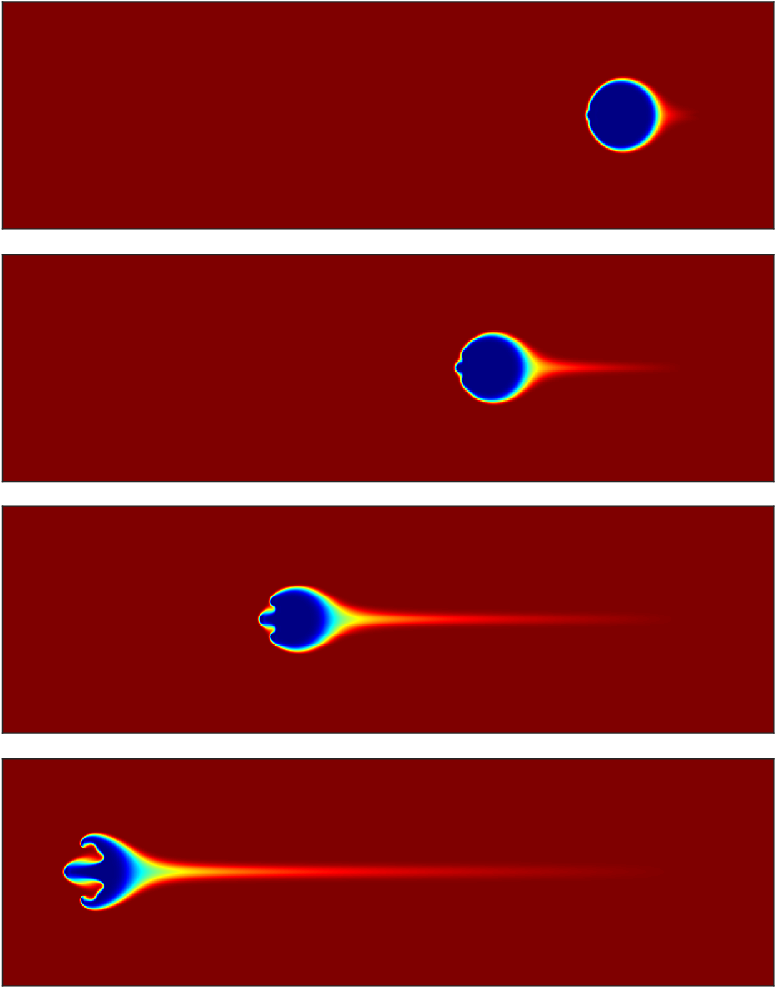}
    \includegraphics[width=0.24\textwidth]{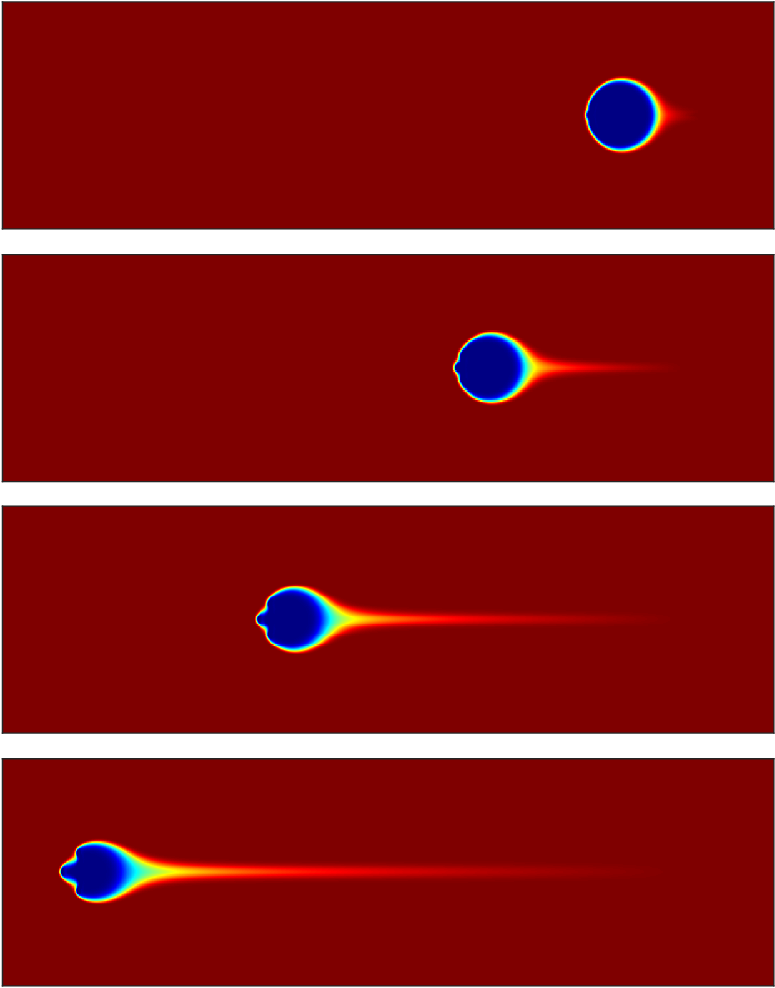}
    \includegraphics[width=0.24\textwidth]{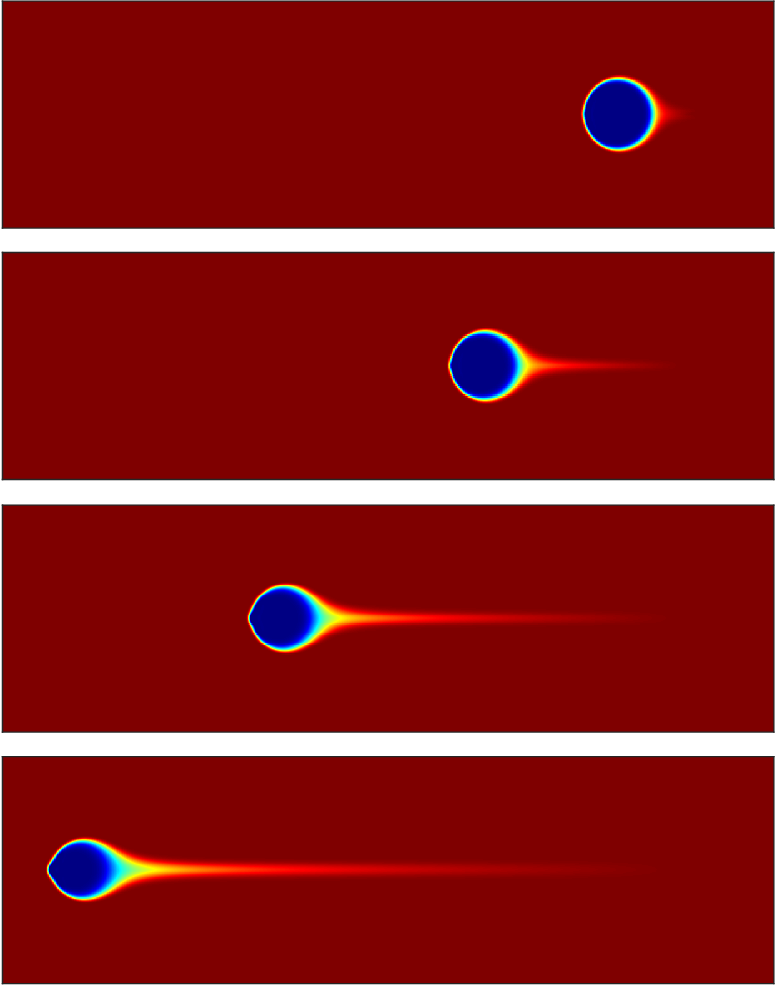}
    \caption{Spatial distribution of a circular blob of radius $r = 0.5$ at different time, $t = 2, 4, 7, 10$ (from top to bottom) for $Pe = 3000$ and and (a) $R = 3$, (b) $R = 3.4$, (c) $R = 3.5$, and (d) $R = 3.9$.}
    \label{fig:Pe3000}
\end{figure}

The resultant flow in the vicinity of the blob is determined by a competition between the penetration of the displacing fluid into the more viscous blob and the displacing fluid flowing past the blob. At a low mobility ratio (small $R$), the displacing fluid easily penetrates the blob. However, due to low velocity gradients, the blob experiences minimal deformation and elongates only in the longitudinal direction, resulting in a comet-shaped deformation with a downstream tail, without undergoing viscous fingering dynamics. An increase in the mobility ratio enhances the velocity gradient at the rear half of the blob, leading to the tendency to form a finger-like pattern, yet the displacing fluid fails to break through the blob. Thus, the blob deforms into a lump-shaped pattern with an upstream nose and a downstream tail. For a given $Pe$, the value of $R$ that corresponds to this transition from comet shape to lump shape is denoted by $R_{C \rightarrow L}$. As $R$ is further increased beyond a threshold value $R_{\rm VF}^l$, the displacing fluid penetrates the blob and deforms the latter into a finger-like structure (e.g., see figure \ref{fig:Pe1500}). This behavior persists for a finite window of $R$ bounded above by $R_{\rm VF}^u$. A further increase in $R$ leads to stronger velocity gradients between the displacing and displaced fluids. As a result, the less viscous fluid flows around the more viscous blob instead of penetrating the latter. Thus, the blob deforms into a lump-shaped pattern for $R \in (R_{\rm VF}^u, R_{L \rightarrow C}]$ (for an example of a lump-shaped deformation, refer to figure \ref{fig:Pe3000}(c)) and a comet-shaped pattern for $R > R_{L \rightarrow C}$ (for examples of comet-shaped deformation, refer to figure \ref{fig:Pe3000}(d) and figure \ref{fig:Pe1000}). Here, $R_{L \rightarrow C}$ denotes the value of $R$ corresponding to the transition from the lump-shaped region to the comet-shaped region. 

The following best fit of the computed lower and upper limits of the interval for viscous fingers provides root-mean-squared (RMS) error less than $0.05$: 
\begin{eqnarray}
    \label{eq:R_l}
    & R_{\rm VF}^l \approx 317.68 e^{-0.007 Pe} + 0.8279 e^{-0.0005 Pe}, \\ 
    \label{eq:R_u}
    & R_{\rm VF}^u \approx 2.1963 e^{0.0001 Pe} - 13.675 e^{-0.0027 Pe}. 
\end{eqnarray}
Similarly, the best fit of the computed values of $R$ and $Pe$ demarcating the transition from comet (lump) shape to lump (comet) shape, with an RMS error less than $0.05$, yield
\begin{eqnarray}
    \label{eq:R_CL}
    & & R_{C \rightarrow L} \approx 2.3919 e^{-0.0021 Pe} + 0.3154 e^{-0.0003 Pe} \qquad (Pe \geq 800), \\ 
    \label{eq:R_LC}
    & & R_{L \rightarrow C} \approx 3.2482 e^{0.0001 Pe} - 5.764 e^{-0.0015 Pe} \qquad (Pe \geq 750). 
\end{eqnarray}
    



\subsection{Spatio-temporal dynamics} \label{subsec:spatio-temporal}

In this section, we discuss the spatio-temporal dynamics of a high-viscosity circular blob at different flow conditions (in terms of the dimensionless parameters, $Pe$ and $R$). 

Figure \ref{fig:Pe3000} depicts deformation of the circular blob ($\bigstar$ markers in figure \ref{fig:Phase_plane}) corresponding to all three regions in the phase plane --- $R = 3, 3.4$ (VF), $3.5$ (lump), and $3.9$ (comet). We noticed that the onset of fingering dynamics for $R = 3.4$ is delayed than $R = 3$ -- a counterintuitive result as compared to the rectilinear displacement of a finite slice \citep[see][]{de2005viscous}. As mentioned in \S \ref{subsec:phase_diagram}, with increasing values of $R$ in the interval $(R_{\rm VF}^l, R_{\rm VF}^u)$, the less viscous displacing fluid flows around the blob more than the former penetrates the latter. This explains the delayed onset for $R = 3.4 (= R_{\rm VF}^u)$ than $R = 3$.
For $R = 3.5 (> R_{\rm VF}^u)$, the less viscous fluid attempts to penetrate the more viscous blob, and a lump-shaped deformation is developed similar to the case of $R = 3.4$ (see panel 2 in \ref{fig:Pe3000}(b) and \ref{fig:Pe3000}(c) at $t = 4$). However, due to the stronger viscous resistance of the blob, it eventually deforms into a lump-shaped pattern. Finally, for $R = 3.9 (\approx R_{L \rightarrow C})$, the blob deforms into a comet. It is worth noting that for this choice of $R$, there is a signature of a lump-shaped structure at the rear interface of the blob. 


\begin{figure}[!htbp]
    \centering
    (a) \hspace{1.9 in} (b) \hspace{1.9 in} (c) \\ 
    \includegraphics[width=0.32\textwidth]{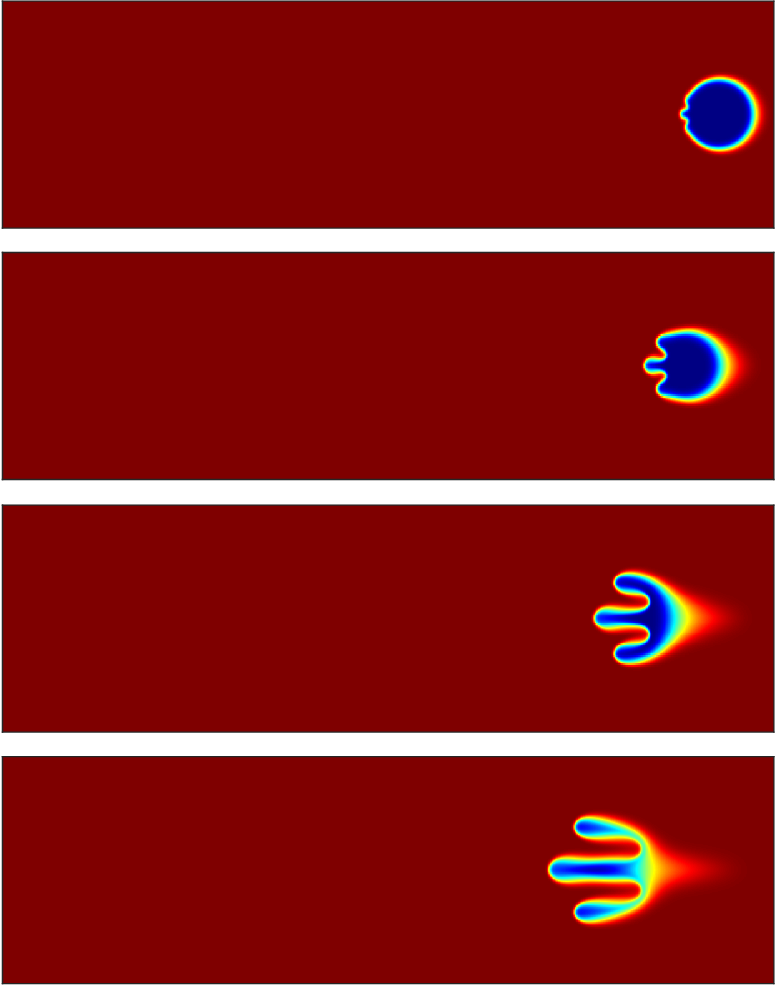}
    \includegraphics[width=0.32\textwidth]{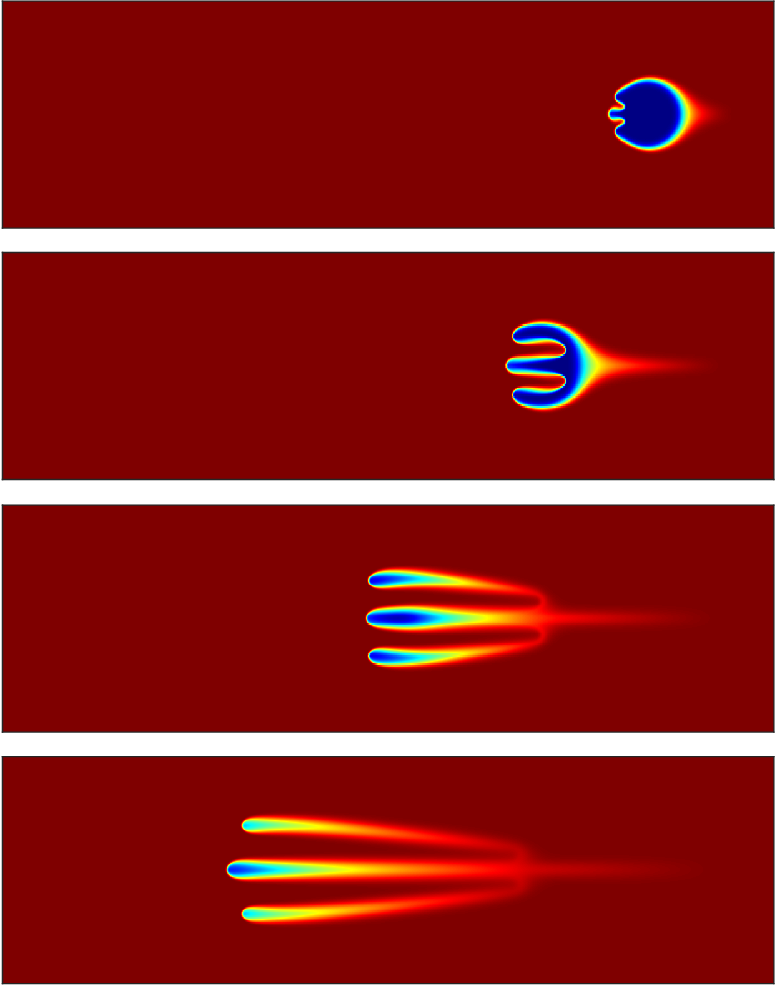}
    \includegraphics[width=0.32\textwidth]{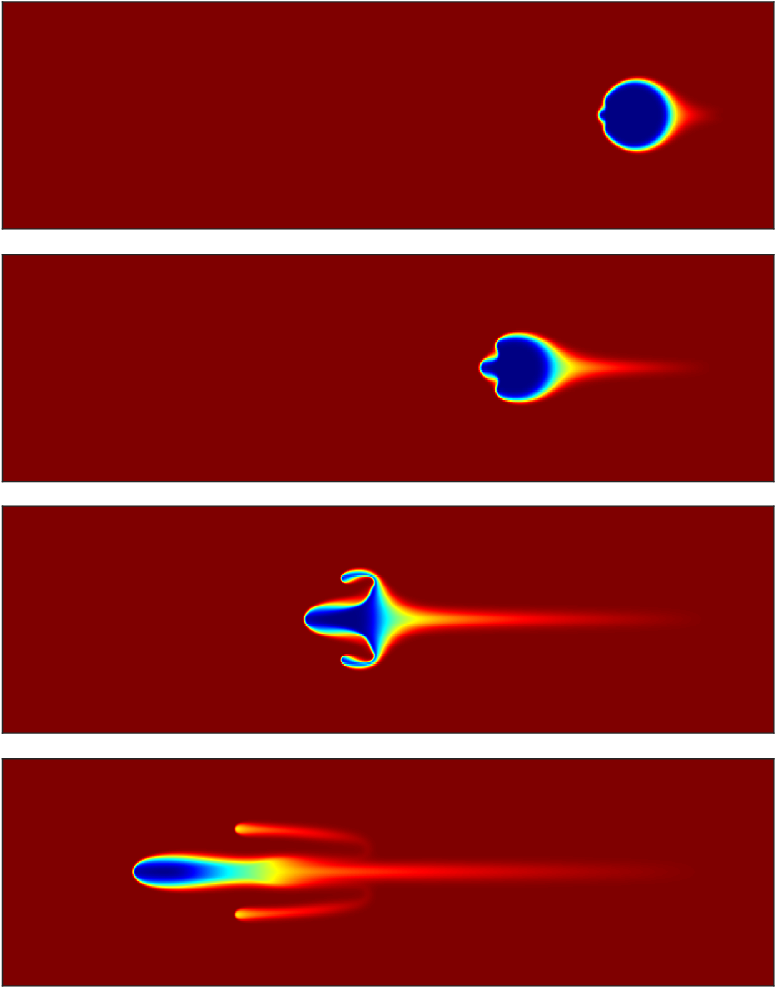}
    \caption{Spatial distribution of a circular blob of radius $r = 0.5$ at $t = 2, 4, 7, 10$ (from top to bottom) for $Pe = 1500$ and (a) $R = 0.5$, (b) $R = 2$, (c) $R = 2.5$.} 
    \label{fig:Pe1500}
\end{figure}


\begin{figure}[!htbp]
    \centering
    (a) \hspace{1.9 in} (b) \hspace{1.9 in} (c) \\ 
    \includegraphics[width=0.32\textwidth]{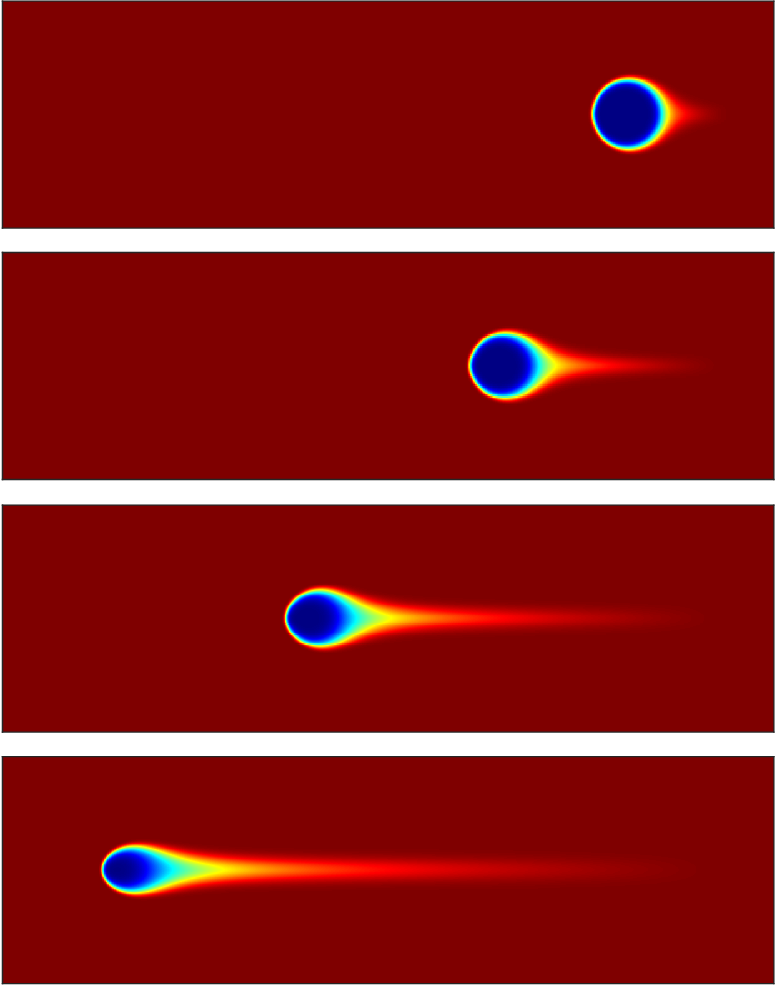}
    \includegraphics[width=0.32\textwidth]{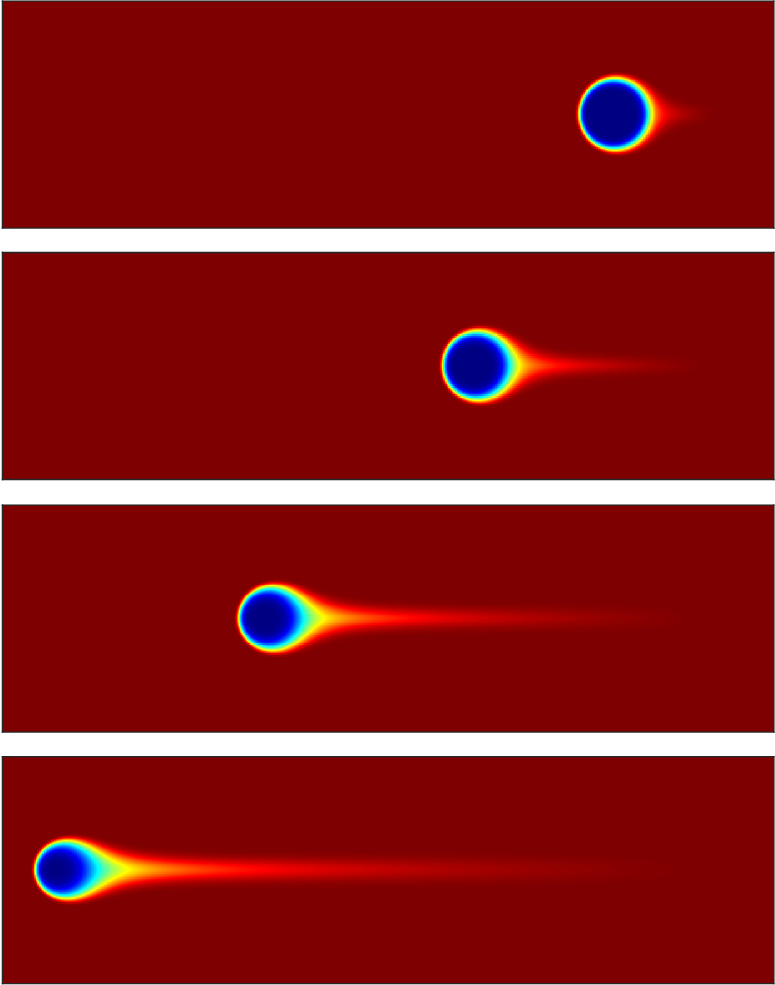}
    \includegraphics[width=0.32\textwidth]{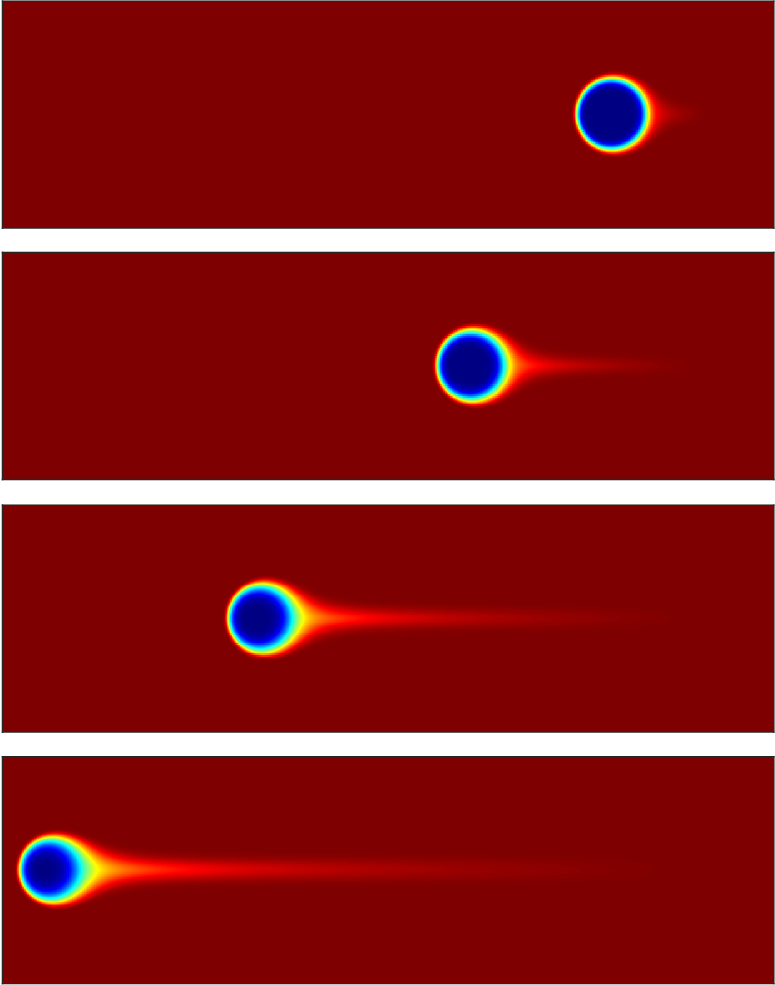} \\
    (d) \\ 
    \includegraphics[width=0.9\linewidth]{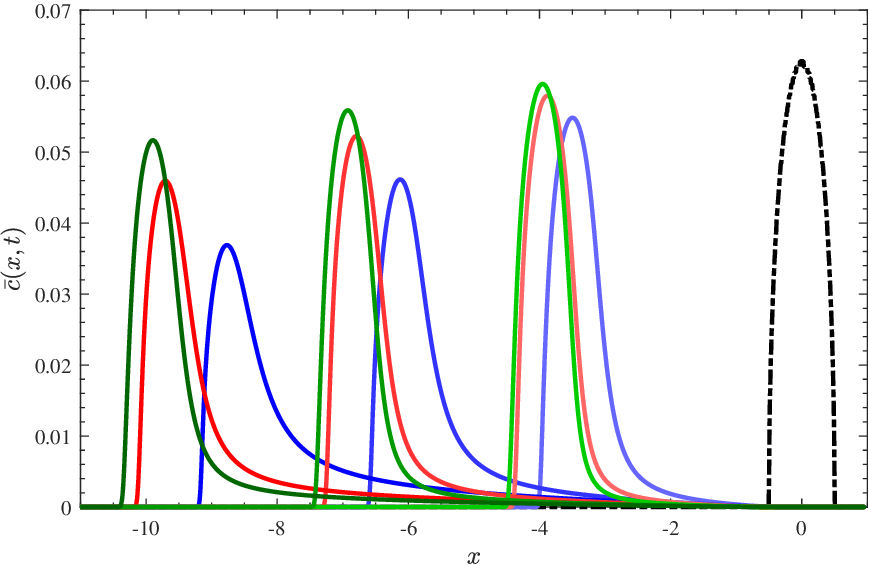}
    \caption{Spatial distribution of a circular blob of radius $r = 0.5$ at different time, $t = 2,  4, 7, 10$ (from top to bottom) for $Pe = 1000$ and (a) $R = 3$, (b) $R = 5$, (c) $R = 7$. Corresponding transverse-averaged concentration profiles are shown in (d) for $R = 3$ (blue), $R = 5$ (red), and $R = 7$ (green) at $t = 4, 7$ and $10$ (light to dark shades). For reference, the initial distribution of the solute is shown in black (dash-dotted line). This clearly depicts a thin-tailed distribution of the solute concentration in the longitudinal direction.}
    \label{fig:Pe1000} 
\end{figure}

Next, we dive deep into the dynamics of viscous fingering and comet-shaped deformation of the blob. Figure \ref{fig:Pe1500} depicts the fingering dynamics for $Pe = 1500$ and $R = 0.5, 2, 2.5$ ($\lozenge$ markers in figure \ref{fig:Phase_plane}). The less viscous ambient fluid penetrates the more viscous blob and deforms the latter into a trident shape. The characteristic features of the fingering dynamics depend on the log-mobility ratio. The blob experiences the least deformation for $R = 0.5$ (see figure \ref{fig:Pe1500}(a)); whereas, the longest tail formation is observed for $R = 2.5$ (see figure \ref{fig:Pe1500}(c)). Interestingly, the onset of viscous fingering is the earliest for $R = 2$ among the three values considered here. As anticipated, and consistent with the flow dynamics highlighted in \S \ref{subsec:phase_diagram},  the onset of viscous fingering in a more viscous blob depends non-trivially on $R$ compared to its finite slice counterpart.

For $Pe = 1000$, the comet-shaped deformation of the blob for $R > R_{L \rightarrow C}$ depicts an increasing tail length as $R$ increases. To further quantify this, we have analyzed the corresponding transverse-averaged concentration, $\bar{c}(x, t)$, portraying a thin-tailed distribution of the solute concentration in the longitudinal direction. In particular, we consider $R = 3$ ($\bigtriangledown$ in figure \ref{fig:Phase_plane}), $5$ and $7$. Spatio-temporal evolutions of the blob capture two key features: (i) the mean position of the sample shifts further upstream, and (ii) the blob spreads more in the transverse direction with increasing viscosity contrast (see figure \ref{fig:Pe1000}(a-c)). With increasing $R$, the displacing fluid minimally penetrates the blob, resulting in a thinner diffusive layer to be carried along with the displacing fluid past the blob. Thus, although the diffusion remains the same for a fixed $Pe$, the blob experiences a larger spreading in the transverse direction with an increasing $R$ beyond $R_{L \rightarrow C}$. These phenomena are also evident from the transverse-averaged concentration (see figure \ref{fig:Pe1000}(d)) depicting the thinnest tail with the highest peak at the furthest upstream position corresponding to $R = 7$. These quantitative features of blob deformation and fingering dynamics are discussed further in \S \ref{subsec:quantitative}. 

\begin{figure}[!htbp]
    \centering
    (a) \hspace{1.3 in} (b) \hspace{1.3 in} (c) \hspace{1.3 in} (d) \\ 
    \includegraphics[width=0.24\textwidth]{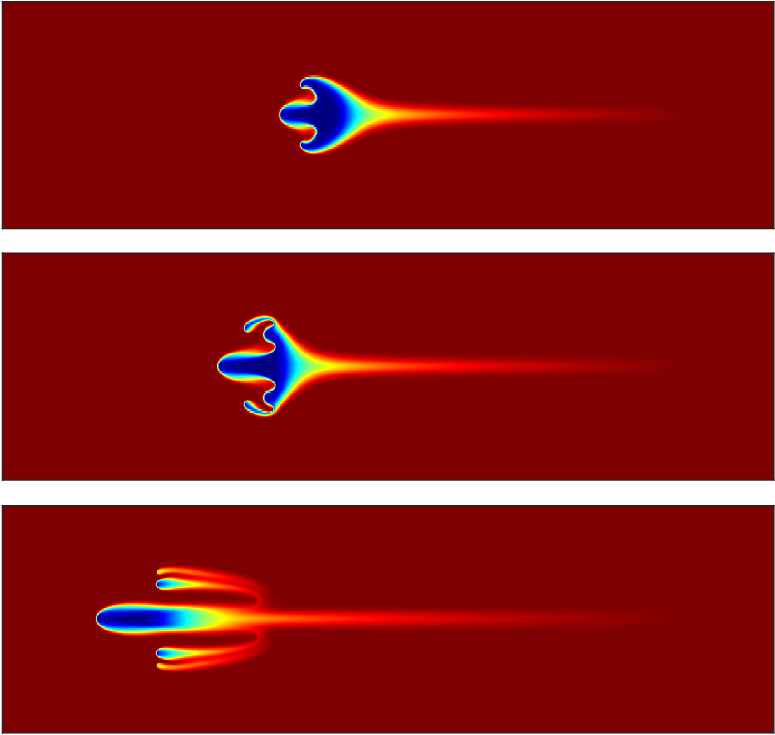}
    \includegraphics[width=0.24\textwidth]{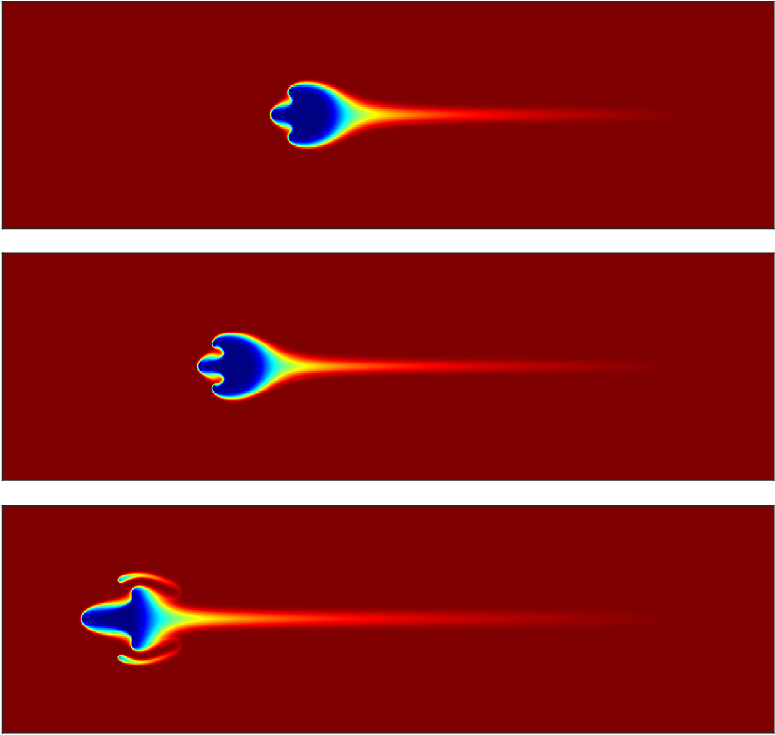}  
    \includegraphics[width=0.24\textwidth]{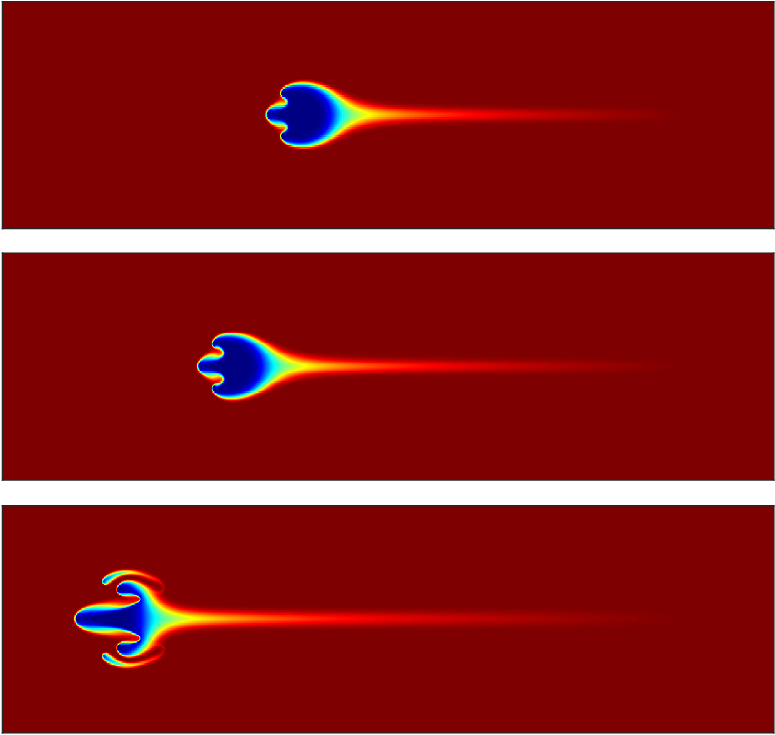} 
    \includegraphics[width=0.24\textwidth]{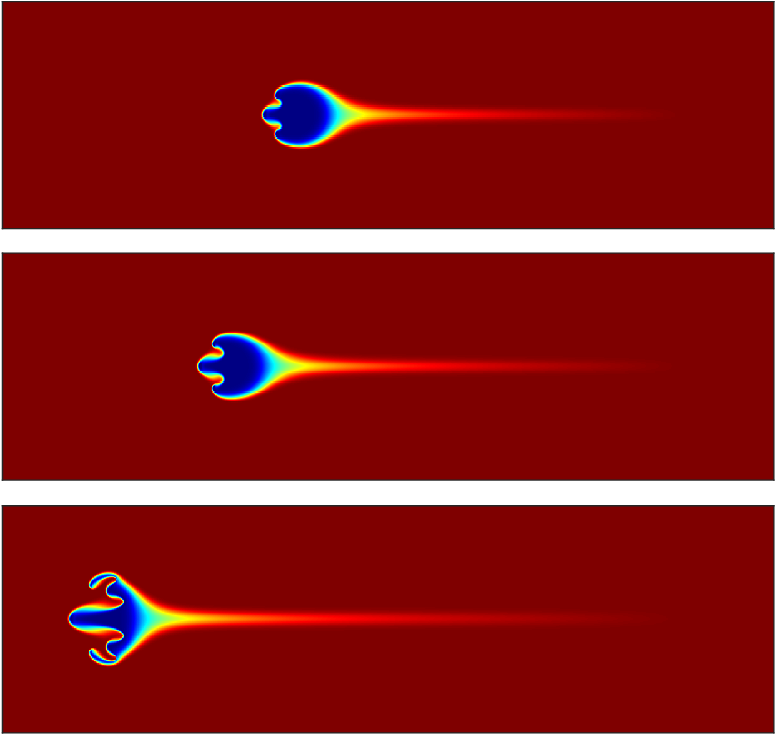}
    \caption{Spatial distributions of a circular blob of radius $r = 0.5$ at $t = 7, 8, 10$ (from top to bottom) for different choices of $Pe$ and $R$ depict finger splitting for ($Pe, R$) = (a) (2000, 2.9), (b) (2250, 3.1), (c) (2500, 3.2), and (d) (2750, 3.3).} 
    \label{fig:splitting}
\end{figure}

We close this section with a discussion of fingering dynamics for specific parameter values ($\triangle$ in figure \ref{fig:Phase_plane}) for which tip-splitting phenomena have been captured (see figure \ref{fig:splitting}). It is observed that the blob initially deformed into a trident-shaped fingering pattern. Subsequently, the finger in the middle moves upstream rapidly compared to the other two adjacent fingers on either side of it, and the latter split into two more fingers, resulting in the formation of five fingers at different instants of the evolution based on the parameter values. It is worth noting that, although some of the parameters reported in figure \ref{fig:splitting} fall within the parameter region explored by \citet{pramanik2015viscous, sharma2017dynamics}, this unique feature was not reported in their studies. Thus, our HOC scheme not only reproduces the fingering dynamics and blob deformation reported in \citet{pramanik2015viscous}, but also unveils new physics while displacing a high-viscosity circular blob by less viscous fluid in homogeneous porous media. 


\subsection{Quantitative effects of $R$ and $Pe$} \label{subsec:quantitative}

Transverse-averaged concentration profile is a classical metric that has been widely utilized in experimental \citep{kretz2003experimental, bacri1991three} as well as in theoretical studies of VF \citep{mishra2008differences, pramanik2016fingering}. This metric helps understand the dynamics averaged along the cross-sectional direction of the domain. Variance of the transverse-averaged concentration defined as \cite{mishra2008differences},
\begin{equation}
    \label{eq:var_x}
    \sigma_x^2(t) = \frac{\int_{0}^{L_x} x^2 \bar{c}(x,t) {\rm d} x}{\int_{0}^{L_x} \bar{c}(x,t) {\rm d} x} - \left[ \frac{\int_{0}^{L_x} x \bar{c}(x,t) {\rm d} x}{\int_{0}^{L_x} \bar{c} (x, t) {\rm d} x} \right]^2, 
\end{equation}
measures the spreading of the solute in the longitudinal direction. Note that the transverse-averaged concentration profile, $\bar{c}(x, t)$, is capable of representing fingering dynamics through a distorted peak; whereas, the variance, $\sigma_x^2(t)$, quantifies variations of the solute concentration in the longitudinal direction. Although these metrics qualitatively and quantitatively represent features of the fingering dynamics of a finite sample [\citet{mishra2008differences} and references therein], localization of the circular blob both in the longitudinal and transverse directions demands further qualitative and quantitative measures. Longitudinal-averaged concentration, $\tilde{c}(y, t)$, and its variance 
\begin{equation}
    \label{eq:var_y}
    \sigma_y^2(t) = \frac{\int_{0}^{L_y} y^2 \tilde{c}(y, t) {\rm d} y}{\int_{0}^{L_y} \tilde{c}(y, t) {\rm d} y} - \left[ \frac{\int_{0}^{L_y} y\tilde{c}(y, t) {\rm d} y}{\int_{0}^{L_y} \tilde{c}(y, t) {\rm d} y} \right]^2.
\end{equation}
capture variations in the transverse direction. 



\begin{figure}[!h]
    \centering
    (a) \hspace{3.2 in} (b) \\ 
    \includegraphics[width=0.495\textwidth]{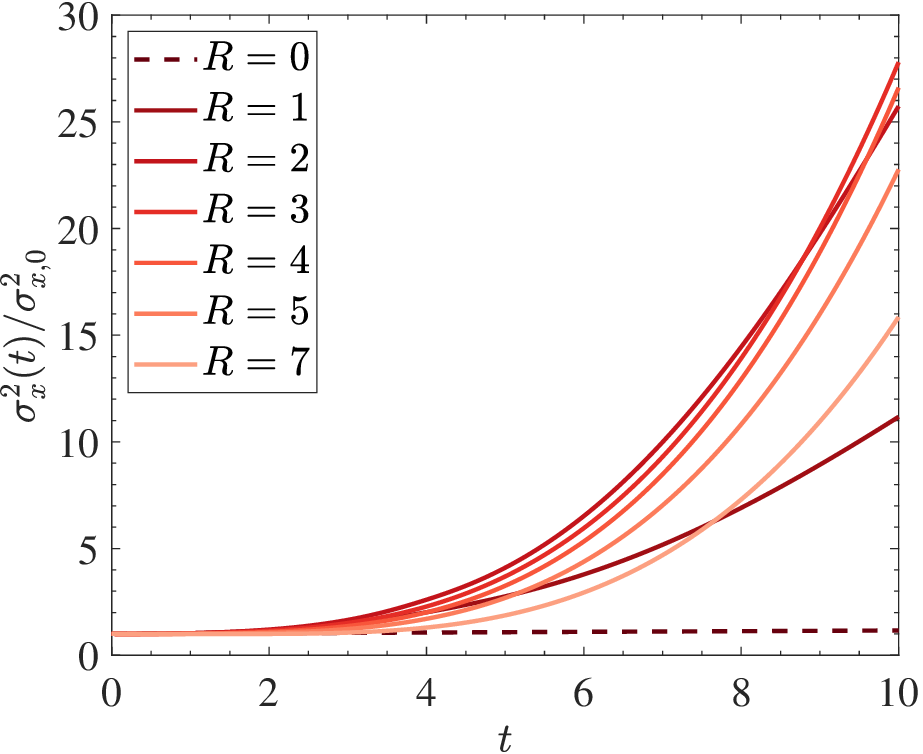} 
    \includegraphics[width=0.495\textwidth]{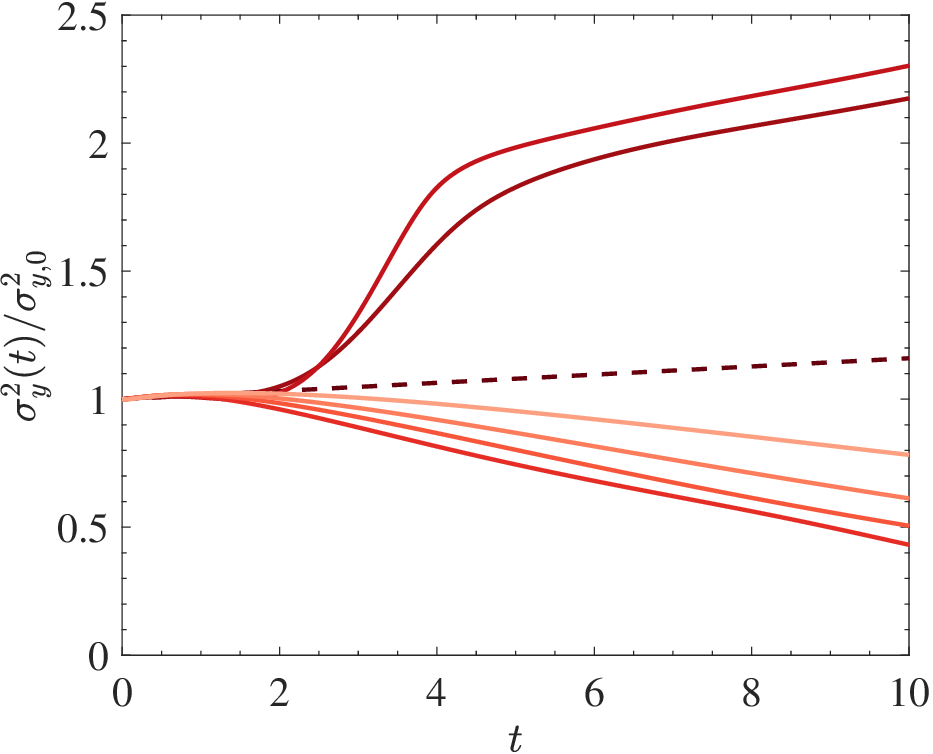} 
    \caption{Effects of log-mobility ratio on the evolution of variances rescaled with their initial value (a) $\sigma_x^2(t)/\sigma_{x, 0}^2$, and (b) $\sigma_y^2(t)/\sigma_{y, 0}^2$ for $r = 0.5$, $Pe = 2000$, where $\sigma_{i, 0}^2 = \sigma_{i}^2(t = 0)$, $i = x, y$. Fingering instability enhances both the longitudinal and transverse spreading ($R = 1, 2$); whereas, lump- and comet-shaped deformations enhance only the longitudinal spreading, but suppress the transverse spreading ($ R \geq 3$).} 
    \label{fig:variance}
\end{figure}

Temporal evolutions of $\sigma_x^2$ and $\sigma_y^2$ for $Pe = 2000$ and different values of $R$ are shown in figures \ref{fig:variance}(a) and \ref{fig:variance}(b), respectively, along with the viscosity-matched case ($R = 0$) shown in a dashed line for reference. Due to radial symmetry for $R = 0$, both $\sigma_x^2$ and $\sigma_y^2$ grow linearly. The initial variations of $\sigma_x^2$ and $\sigma_y^2$ are identical for all $R \geq 0$, indicating that the evolution is dominated by diffusion before the blob deformation. As observed and reported in the spatio-temporal dynamics of the blob in \S \ref{subsec:spatio-temporal}, the tail length increases with $R$. Thus, it is anticipated that at any given instant $\sigma_x^2$ increases with $R$. Contrary to that, figure \ref{fig:variance}(a) reveals that $\sigma_x^2$ varies non-trivially with $R$, attributed to the nonlinear dynamics of the blob. Similar to $\sigma_x^2$, $\sigma_y^2$ also exhibits non-trivial dependence on $R$. Despite these qualitative similarities between the variances of longitudinal- and transverse-averaged concentration, a key dissimilarity between these two metrics highlights the importance of longitudinal-averaged concentration and the corresponding variance. Viscous fingering deformation widens the blob in the transverse direction; whereas, lump- and comet-shaped deformations shrink the blob in this direction. This results in larger (smaller) $\sigma_y^2(t)$ for $R$ corresponding to VF (lump- and comet-shaped deformation) as compared to $R = 0$. Interestingly, as $R$ increases beyond $R_{L \rightarrow C}$, the transverse spreading of the blob is almost similar to that of $R = 0$ (see figure \ref{fig:variance}(b)), barring a thin tail formation -- consistent with the discussion of figure \ref{fig:Pe1000} in \S \ref{subsec:spatio-temporal}. 




\begin{figure}[!h]
\centering
(a) \hspace{3.2 in} (b) \\ 
\includegraphics[width=0.495\linewidth]{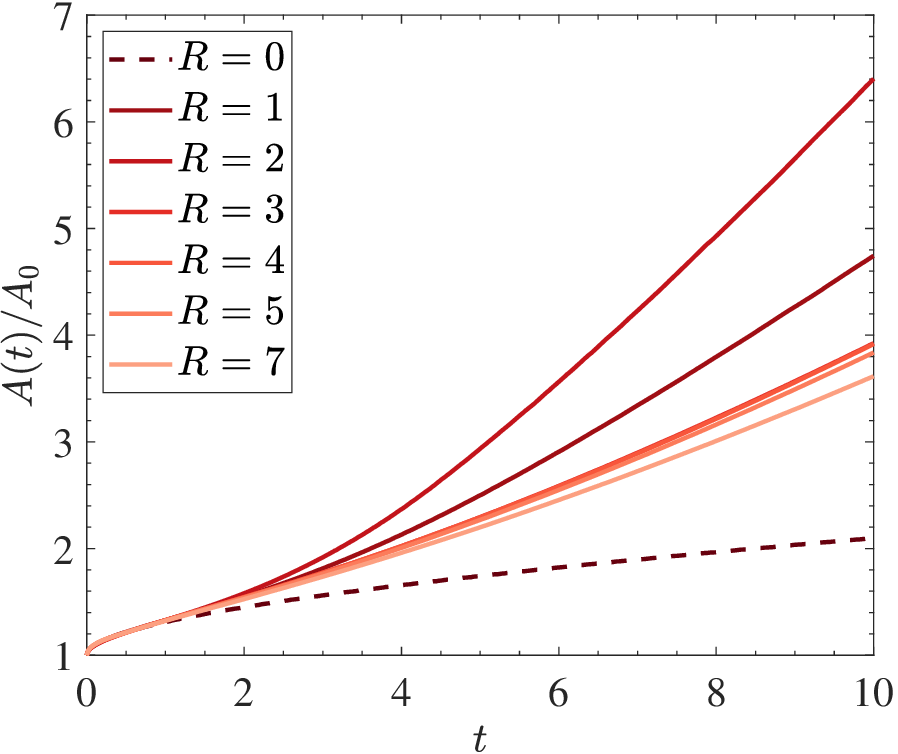} 
\includegraphics[width=0.495\textwidth]{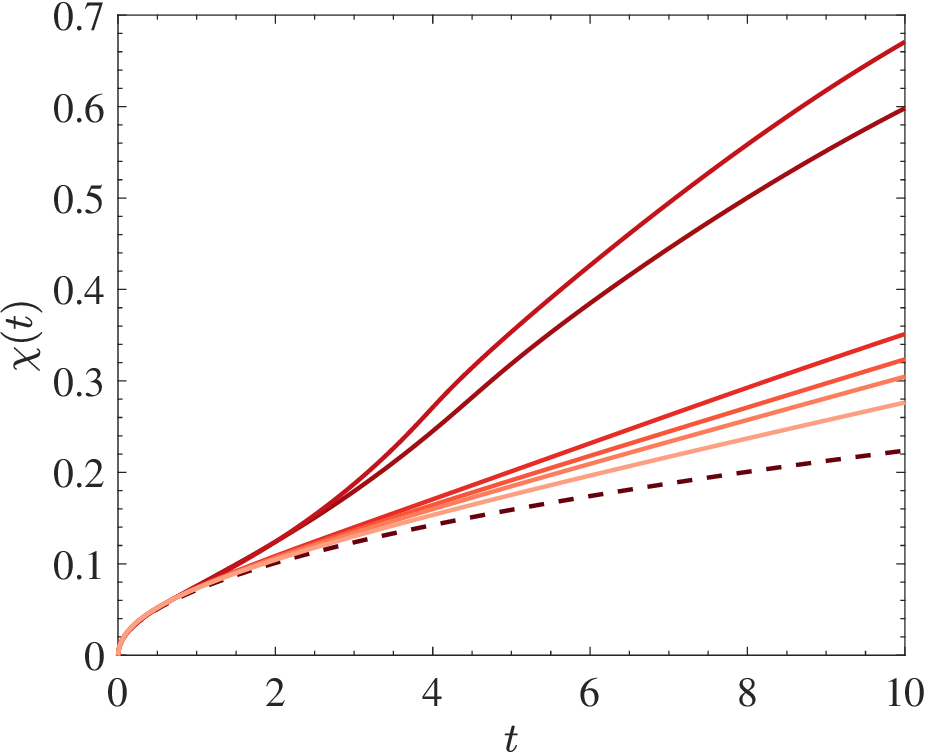}
\caption{Effects of log-mobility ratio on the evolution of (a) area covered by the sample rescaled by its initial value ($A(t)/A_0$) and (b) degree of mixing ($\chi(t)$) for $r = 0.5$ and $Pe = 2000$. Viscous fingering ($R = 1, 2$) mixes the blob more rapidly as compared to the lump- and comet-shaped deformations ($R \geq 3$).} 
\label{fig:degree_mixing}
\end{figure}

The variances of the longitudinal- and transverse-averaged concentration help gain insight into the spreading of the blob in the transverse and longitudinal directions, respectively. However, they fail to comprehensively capture the area over which the blob spreads due to the deformation. In this direction, we measure 
\begin{eqnarray}
    \label{eq:area}
    A(t) = \iint_{\Omega(t)} {\rm d} x {\rm d} y, 
\end{eqnarray}
where $\Omega(t)$ is defined as, $\Omega(t) = \left\{ (x,y): c(x,y,t) \geq 0.01 \right\}$. As anticipated, VF results in larger $A(t)$ as compared to the lump- and comet-shaped deformations. Temporal evolution of the area covered by the miscible circular blob depicts that the blob spreads over the largest area for $R = 2$ among the parameters explored in figure \ref{fig:degree_mixing}(a). In summary, as compared to the viscosity-matched case ($R = 0$), (i) a high viscous blob ($R > 0$) spreads more in the longitudinal direction due to tail formation irrespective of the nature of deformation, (ii) transverse spreading of the blob depends on the nature of deformation, and (iii) the area covered by the blob is larger. 

Finally, we quantify the mixing of the blob with the ambient displacing fluid using \citep{pramanik2016fingering} 
\begin{eqnarray}
    \label{eq:mixing}
    \chi(t) = 1 - \frac{\sigma^2(t)}{\sigma^2_{\rm max}}. 
\end{eqnarray}
Here, $\sigma^2(t) = \langle c^2 \rangle - \langle c \rangle^2$ is the global variance of the solute concentration, and $\langle \cdot \rangle$ denotes the spatial average over the domain, i.e.,
\begin{equation}
    \label{eq:spatial_avg} 
    \langle c \rangle (t) = \frac{1}{L_x L_y} \int \limits_{0}^{L_x} \int \limits_{0}^{L_y} c(x, y, t) {\rm d} x {\rm d} y. 
\end{equation}
The maximum variance, $\sigma^2_{\rm max}$, results in $\chi = 0$, indicating a perfectly segregated state; whereas, $\sigma^2 = 0$ corresponds to $\chi = 1$, which represents a perfectly mixed state. Figure \ref{fig:degree_mixing}(b) captures the temporal evolution of global variance as well as the degree of mixing of the miscible circular blob. Mixing of the sample is a result of diffusive spreading, viscous fingering, and the downstream tail formation. It is observed that in the initial diffusion-dominated regime, $\chi(t)$ remains almost identical for all values of $R \geq 0$. At a later stage, when the advection dominates, mixing enhances rapidly due to fingering deformation as compared to lump- and comet-shaped deformations (compare curves corresponding to $R = 1, 2$ with the curves corresponding to $R \geq 3$ in figure \ref{fig:degree_mixing}(b)). 





\begin{figure}[!h]
\centering
(a) \hspace{3.2 in} (b) \\ 
\includegraphics[width=0.495\textwidth]{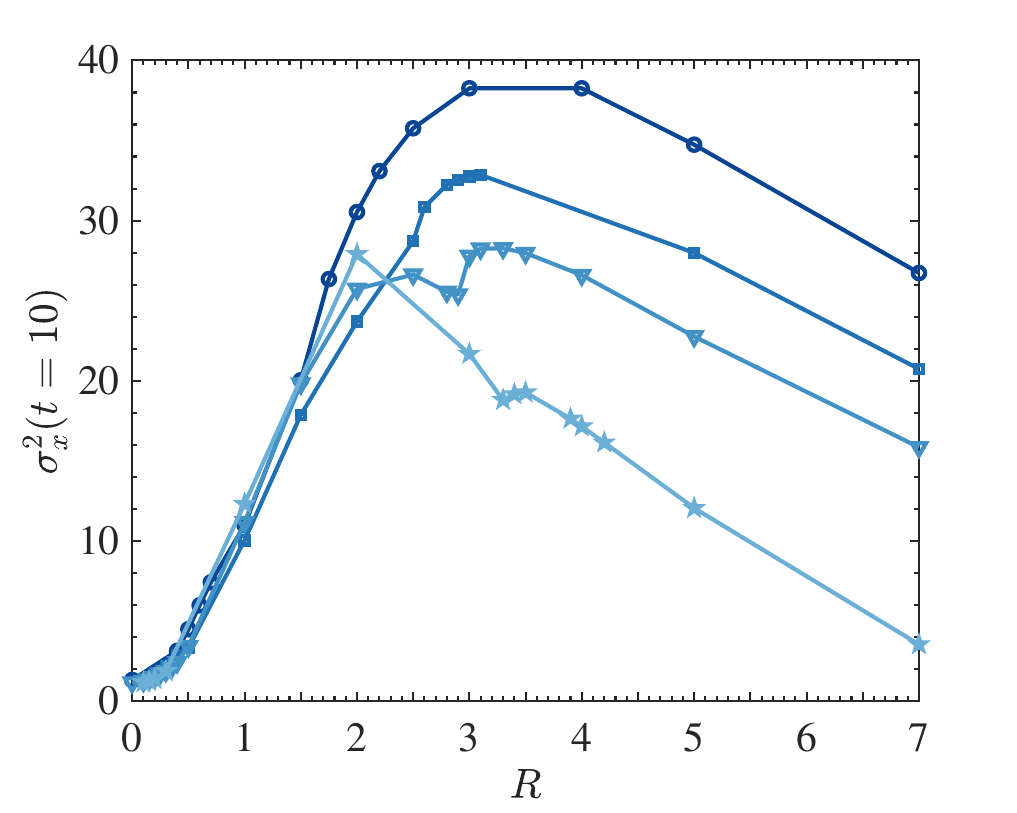}
\includegraphics[width=0.495\textwidth]{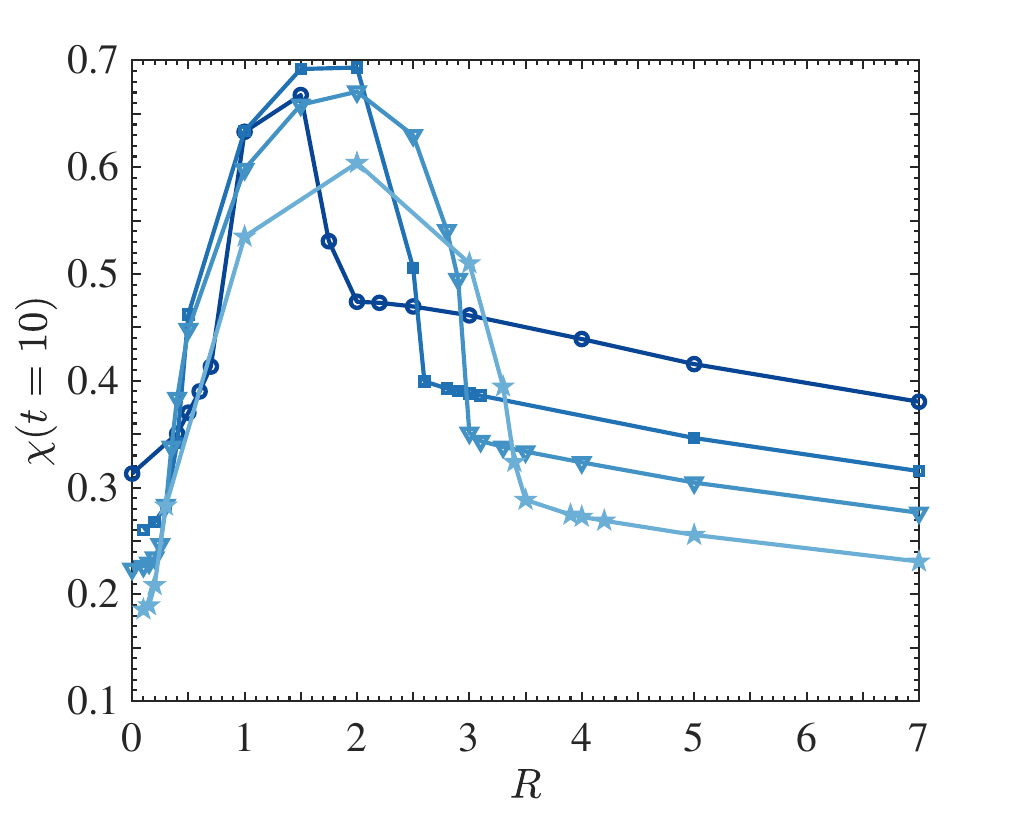}
\caption{Variation of (a) $\sigma_x^2(t = 10)$ and (b) $\chi(t = 10)$ with $R$ for $Pe = 1000$, $1500$, $2000$, and $3000$ (dark to light) exhibit that the the maximal variance of the blob in the longitudinal direction and the maximal mixing are attained for an intermediate value of $R$. }
\label{fig:optimal_mixing}
\end{figure}

We summarize the effects of log-mobility ratio on the longitudinal spreading and mixing of a blob in figure \ref{fig:optimal_mixing}. We note that the $\sigma_x^2$ and $\chi$ at the end of the simulation ($t = 10$), i.e., $\sigma_x^2(t = 10)$ and $\chi(t = 10)$ vary non-monotonically with $R$. Since the mixing enhances at due to fingering instability, $\chi(t = 10)$ rapidly changes for $R \in (R_{\rm VF}^l, R_{\rm VF}^u)$ and the maximum mixing happens for $R \approx (R_{\rm VF}^l + R_{\rm VF}^u)/2$. On the other hand, $\chi(t = 10)$ does not experience significant changes for $R > R_{VF}^u$. Since the longitudinal spreading ($\sigma_x^2$) bears the signature of the tail as well as the leading finger in the upstream direction, $\sigma_x^2(t = 10)$ does not attain its maximum value for $R \approx (R_{\rm VF}^l + R_{\rm VF}^u)/2$. 

\section{Conclusion} \label{sec:conclusion}

In this work, we have numerically studied the deformation of a more viscous circular blob displaced rectilinearly by a less viscous fluid in a homogeneous porous medium. A temporally second-order and spatially fourth-order accurate finite difference scheme on a compact uniform grid has been used for numerical simulations. Contrary to the periodic boundary conditions used in some earlier studies, the current study employed zero-flux boundary conditions for the concentration; whereas, for streamfunction zero-flux and Dirichlet conditions are used in the longitudinal and transverse directions, respectively. The stability of the rear interface of the blob has been studied on a wide range of P\'eclet number and log-mobility ratio. Moreover, the use of the HOC scheme in the computation allowed the simulation for $Pe \leq 3000$ and $R \leq 7$. Our analysis confirms the existence of a finite critical window of $R \in (R_{\rm VF}^l, R_{\rm VF}^u)$ for viscous fingering instability. The best fit of the computed values of $R_{\rm VF}^l(Pe)$ and $R_{\rm VF}^u(Pe)$ reveals that these critical values follow $a e^{b \; Pe } + c e^{d \; Pe}$ relation for some suitably chosen constants $a$, $b$, $c$ and $d$. These best fit can be used as indicators to approximate critical windows for fingering instabilities at $Pe > 3000$. Quantitative analyses of longitudinal and transverse spreading of the blobs and their mixing in the ambient fluid exhibit non-monotonic dependence on $R$. An optimal mixing of the blob depends on the competition between the fingering instability and tail formation. These findings indicate that an optimal mixing (vs. spreading) of a high viscous circular blob in porous media can be achieved by suitably controlling the flow parameters. 

\section*{Acknowledgments}

Authors acknowledge financial support through the Core Research Grant (CRG/2023/004156) supported by the Science and Engineering Research Board, Department of Science and Technology, Government of India. 
S.P. acknowledges financial support through the Start-Up Research Grant (SRG/2021/001269), MATRICS Grant (MTR/2022/000493) from the Science and Engineering Research Board, Department of Science and Technology, Government of India, and Start-up Research Grant (MATHSUGIITG01371SATP002), IIT Guwahati. 



\bibliography{mijanurbib}

\end{document}